\documentclass[aps]{revtex4-2}

\usepackage{eurosym}
\usepackage{graphicx}
\usepackage[colorlinks=true, pdfstartview=FitV, linkcolor=blue, citecolor=red, urlcolor=magenta]{hyperref}
\usepackage{graphicx}
\usepackage{latexsym}
\usepackage{amsmath}
\usepackage{amsfonts}
\usepackage{amssymb}
\usepackage{verbatim}
\usepackage{braket}
\usepackage{extarrows}
\usepackage{hyperref}
\usepackage{orcidlink}
\usepackage{geometry}
\usepackage[english]{babel} 
\usepackage{tikz}
\usetikzlibrary{arrows.meta, decorations.pathmorphing, 3d,calc}

\begin{document}

\title{Vacuum-induced current density from a magnetic flux threading a cosmic dispiration in $(D+1)$-dimensional spacetime}

\author{Herondy~Mota~\orcidlink{0000-0002-7470-1550}}
\email{hmota@fisica.ufpb.br}





\affiliation{Departamento de F\'{\i}sica, Universidade Federal da Para\'{\i}ba,\\
Caixa Postal 5008, Jo\~{a}o Pessoa, Para\'{\i}ba, Brazil}

\begin{abstract}

We investigate the vacuum-induced current density for a charged scalar field in a $(D+1)$-dimensional cosmic dispiration spacetime threaded by a magnetic flux. This background combines a cosmic string and a screw dislocation, yielding a nontrivial helical geometry. By constructing the normalized mode functions of the Klein--Gordon equation, we evaluate the Wightman function and obtain the vacuum expectation value of the current density. 
We show that, in addition to the azimuthal component describing a persistent current around the defect, a nonvanishing axial component is induced as a direct consequence of the helical structure of the spacetime. Both components are periodic functions of the magnetic flux, depending only on its fractional part, reflecting the Aharonov--Bohm nature of the effect. Closed expressions are obtained for massive and massless fields in arbitrary dimensions.
We demonstrate that the screw dislocation parameter plays a crucial role in the behavior of the induced currents, leading to the regularization of the components at the origin and controlling their magnitude. Their asymptotic behavior is also analyzed in detail. Our results reduce to known expressions in the absence of the screw dislocation, providing a consistency check. In particular, we examine the physically relevant $(3+1)$-dimensional case, where numerical analysis reveals nontrivial features arising from the interplay between topology and gauge effects.
\end{abstract}

\maketitle

\section{Introduction}
\label{intro}

Quantum field theory in nontrivial backgrounds provides a powerful framework to investigate how geome\-try and topology influence vacuum fluctuations and, consequently, observable physical quantities. It is well established that, even in locally flat spacetimes, boundary conditions and spacetime topology can induce nontrivial vacuum expectation values (VEVs) of physical observables. This phenomenon underlies a variety of effects, including the Casimir effect and vacuum polarization in the presence of gravitational or gauge fields~\cite{bordag2009advances, mostepanenko1997casimir, elizalde1995zeta, milton2001casimir}.

Among the most extensively studied configurations in this context are spacetimes containing linear topo\-logical defects, such as cosmic strings~\cite{Vilenkin1994, Hindmarsh:1994re}. These objects, which may have been formed during phase transitions in the early universe, are characterized by a conical topology that modifies the spectrum of quantum fluctuations. As a consequence, in an idealized zero-thickness cosmic string spacetime, vacuum expectation values of quantities such as the energy-momentum tensor and current densities acquire nonzero contributions, even though the spacetime is locally Minkowskian~\cite{Braganca:2019mvj, Mota:2017slg, DeLorenci:2002jv, deFarias:2020xms, Braganca:2014qma, BezerradeMello:2014phm, DeLorenci:2002jv, BezerradeMello:2011sm, Saharian2007}.

Vacuum polarization effects in cosmic string spacetimes have been extensively investigated in the literature. Most analyses employ the idealized model of a zero-thickness cosmic string, which captures the nontrivial topology generated by the planar angle deficit while allowing for exact analytical treatments. However, more realistic descriptions based on strings with internal structure have also been considered. Examples include finite-core cosmic strings, Nielsen-Olesen vortices and magnetic flux tubes of finite radius, where the internal structure of the defect may give rise to additional vacuum polarization effects beyond those associated with the exterior geometry alone~\cite{Khusnutdinov:1998tf, Khusnutdinov:2004ux, BezerradeMello:2014phm, Koike:2024vhd, Graham:2025udo, Bordag:2001qi}. 

Different observables have been analyzed in these backgrounds, including the vacuum expectation values of the field squared, the energy-momentum tensor and induced vacuum currents. For local quantities such as $\langle \phi^2 \rangle$ and $\langle T_{\mu\nu}\rangle$, the ultraviolet divergences are of local origin and are removed through the standard renormalization procedure, whereas finite-core models may additionally contribute finite structure-dependent terms. Induced vacuum currents exhibit a qualitatively different behavior: the Minkowski contribution vanishes identically, and the resulting currents are determined entirely by the nontrivial topology and gauge structure of the background, as we will see in the present work.

A natural generalization of the cosmic string geometry is provided by the so-called `cosmic dispiration', which consists of a combined topological defect incorporating both a conical structure, which can also be a disclination, and a helical distortion in the form of a screw dislocation~(see Ref.~\cite{Mota:2017slg, DeLorenci:2002jv} and references therein). This spacetime arises, for instance, in the framework of Einstein–Cartan theory and can also be interpreted as an effective geometry in condensed matter systems with line defects. The presence of torsion, associated with the screw dislocation, introduces qualitatively new features when compared to the pure cosmic string case, leading to richer vacuum polarization effects~\cite{Mota:2017slg, DeLorenci:2002jv}.

Quantum field theoretical investigations in dispiration backgrounds have shown that the nontrivial global structure of the spacetime induces modifications in the two-point functions and, consequently, in the VEVs of physical observables. In particular, the renormalized propagator in such geometries reveals that the vacuum becomes polarized, giving rise to nonvanishing components of the energy-momentum tensor~\cite{Mota:2017slg, DeLorenci:2002jv, DeLorenci:2002jv}.

Another important ingredient that can significantly affect vacuum properties is the presence of gauge fields. In particular, a magnetic flux threading the core of a linear topological defect modifies the phase of the quantum modes, leading to observable effects even in regions where the corresponding field strength vanishes. This mechanism is directly related to the Aharonov--Bohm effect and manifests itself, in quantum field theory, through the induction of vacuum currents~\cite{Braganca:2014qma, BezerradeMello:2014phm}.

In cosmic string spacetimes, the presence of a magnetic flux is known to induce azimuthal vacuum currents for charged scalar and fermionic fields~\cite{Braganca:2014qma, BezerradeMello:2014phm, BezerradeMello:2013ohh, MaiordeSousa:2015fvd, SoutoMaiordeSousa:2018pwg, Mohammadi:2014lwa}. These currents depend periodically on the ratio between the magnetic flux and the quantum flux, reflecting the topological nature of the effect. Extensions of this analysis to higher-dimensional and compactified settings have revealed additional contributions, such as axial currents, arising from the interplay between topology and boundary conditions~\cite{Braganca:2014qma}.

In this work, we investigate the vacuum-induced current density associated with a charged scalar field in a $(D+1)$-dimensional cosmic dispiration spacetime in the presence of a magnetic flux running along the defect core. The combined effects of the conical topology, the screw dislocation, and the gauge field lead to a nontrivial modification of the mode structure, which in turn affects the VEV of the current density.

The investigation of VEVs of induced current densities is of particular relevance in quantum field theory in nontrivial backgrounds. In semiclassical approaches, the VEV of the four-current density acts as a source term in Maxwell’s equations, allowing one to assess how quantum fluctuations modify the dynamics of the electromagnetic field. In this sense, the induced current encodes information about vacuum polarization effects and provides a direct link between the underlying quantum field and observable electromagnetic responses.

Moreover, induced currents are especially sensitive to global and topological properties of the background geometry and gauge configuration. Even in locally flat spacetimes, nontrivial topology or the presence of gauge fluxes can lead to nonvanishing VEVs, reflecting the nonlocal structure of the quantum vacuum. This makes the current density a powerful probe of Aharonov--Bohm-type effects, where physical observables depend on gauge potentials despite the absence of local field strengths.

From a more fundamental perspective, the study of induced currents contributes to a deeper understanding of how boundary conditions and topological defects affect the spectrum of quantum fluctuations. These effects are closely related to vacuum polarization phenomena and Casimir-type energies, and may have implications in both high-energy physics and condensed matter analog systems.

Finally, in scenarios involving defects such as screw dislocations, cosmic strings or their analogues (disclinations), the induced current density provides insight into how quantum fields respond to the spacetime topologies and magnetic fluxes, revealing characteristic periodicities that are ultimately rooted in the topo\-logy of the system.

Our analysis is based on the construction of a complete set of normalized mode functions for the Klein--Gordon equation in this background. From these solutions, we evaluate the positive-frequency Wightman function and use it to compute the induced current density. Particular attention is devoted to identifying how the geometric parameters associated with the defect and the magnetic flux influence the behavior of the vacuum current.

The paper is organized as follows. In Sec.~\ref{secIII} we introduce the spacetime geometry of the cosmic dispiration and derive the corresponding solutions of the Klein--Gordon equation. In Sec.~\ref{secIV} we construct the Wightman function and evaluate the vacuum-induced current density. Finally, in Sec.~\ref{secV} we summarize our main results and discuss their physical implications.

Through this paper we use natural units in which both the
Planck constant and the speed of light are set equal to one, $\hbar=c=1$.

\section{The spacetime geometry and Klein-Gordon equation}
\label{secIII}
In this section, we focus on a $(D+1)$-dimensional cosmic dispiration background whose global structure is determined by the simultaneous presence of a conical defect, like a cosmic string, and a helical distortion, characterizing a screw dislocation. The scalar field dynamics is described by the Klein-Gordon equation, whose mode solutions are constructed by imposing regularity at the symmetry axis. Within this setting, we analyze how the geometric and topological properties of the background manifest in the VEVs of physical quantities, with particular emphasis on the induced current density. 

We begin by considering the line element describing a $(D+1)$-dimensional cosmic dispiration spacetime, written in generalized cylindrical coordinates as~\cite{Mota:2017slg}
\begin{equation}
ds^2 = dt^2 - dr^2 - r^2 d\phi^2 - (dz + \kappa \, d\phi)^2 - \sum_{i=4}^{D} (dx^i)^2.
\label{line_E1}
\end{equation}
Here, $\kappa$ is a constant parameter that characterizes the screw dislocation. The coordinates $(t, r, \phi, z, x^4, \ldots, x^D)$ span the ranges
\begin{eqnarray}
r \geq 0, \quad 0 \leq \phi \leq \phi_0 = \frac{2\pi}{q}, \quad -\infty < (t, z, x^i) < +\infty, \quad i = 4, \ldots, D.
\label{rangeC}
\end{eqnarray}

The parameter $q \geq 1$ encodes the presence of the cosmic string, which is assumed to lie along the $(D-2)$-dimensional hypersurface defined by $r=0$. In the particular case $D=3$, this parameter is related to the linear mass density $\mu_0$ of the string through the relation
\begin{eqnarray}
q^{-1} = 1 - 4 G \mu_0,
\end{eqnarray}
where $G$ denotes Newton's gravitational constant.

\begin{figure}[h]
\centering
\begin{tikzpicture}[scale=1.3, line join=round, line cap=round, >=Latex,
                    x={(0.866cm,0.5cm)}, y={(-0.866cm,0.5cm)}, z={(0cm,1cm)}]

\def\R{1.7}
\def\H{4}

\draw[black!70!black] (0,0,-0.2) circle (0.12);

\draw[, thick] (0,0,-0.2) -- (0,0,\H+0.9) node[above] {$z$};

\draw[gray!70] (\R,0,0) arc (0:360:\R);
\draw[gray!70] (0,0,\H) -- (\R,0,0);
\draw[gray!70] (0,0,\H) -- (-\R,0,0);
\draw[gray!70] (0,0,\H) -- (0,\R,0);
\draw[gray!70] (0,0,\H) -- (0,-\R,0);

\draw[very thick, black!70!black] (0.12,0,-0.2) -- (0.12,0,\H+0.3);
\draw[very thick, black!70!black] (-0.12,0,-0.2) -- (-0.12,0,\H+0.3);

\draw[black!70!black] (0,0,\H+0.3) circle (0.12);

\draw[red, thick, domain=0:720, samples=120, smooth, variable=\t]
plot ({0.1*cos(\t)}, {0.1*sin(\t)}, {\H*\t/720});

\draw[->, very thick, blue!70!black] (0,0,-0.2) -- (0,0,\H+0.7);
\node[blue!70!black, right] at (0.15,0,\H/0.9) {$\mathbf{B}$};
%

\def\pitch{1.0}


\draw[<->, thick] (0.3,0,-0.1) -- (0.3,0,0.05+\pitch)
node[midway,right] {$p$};

\end{tikzpicture}

\caption{Schematic representation of a cosmic dispiration with negligible
core structure. The conical structure represents the angular deficit of the
cosmic string, giving rise to a nontrivial topology. A thin magnetic flux tube runs along the defect core, while
the red helix illustrates the screw-dislocation component of the spacetime.
The vertical separation $p=2\pi\kappa/q$ corresponds to the helical pitch.}

\label{fig_dispiration}
\end{figure}
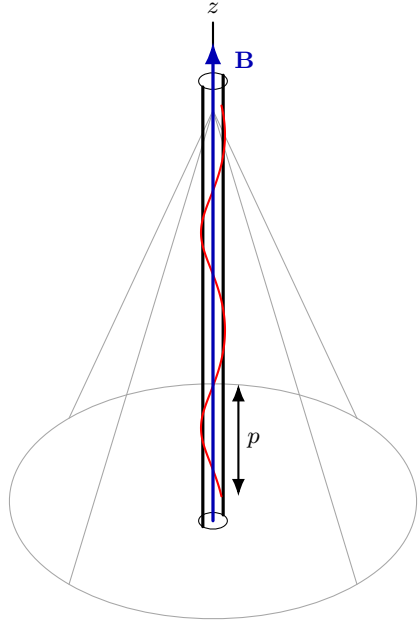

The nontrivial topology of the spacetime can also be implemented by considering the identification condition
\begin{eqnarray}
(r, \phi, Z, x^4, \ldots, x^D) \sim (r, \phi + \phi_0, Z + p, x^4, \ldots, x^D),
\label{IC}
\end{eqnarray}
in a locally flat spacetime given by
\begin{eqnarray}
d s^2=d t^2-d r^2-r^2 d \phi^2-d Z^2-\sum_{i=1}^D\left(d x^i\right)^2,
\label{line_E2}
\end{eqnarray}
which is obtained from Eq.~\eqref{line_E1} by adopting the new spatial coordinate $Z=z+\kappa \phi$. The parameter $p = 2\pi\kappa/q$ is the helical pitch of the cosmic dispiration, as illustrated in Fig.~\ref{fig_dispiration}, and $\phi_0$ is defined in Eq.~\eqref{rangeC}.

Note that, even though the physically relevant case corresponds to $D=3$, we formulate the problem in a more general setting with $D \geq 3$. This extension allows us to track how the resulting quantities are influenced by the presence of additional spatial dimensions. Nevertheless, particular attention will be devoted to the standard $(3+1)$-dimensional spacetime, which represents the main physical scenario of interest.

We consider a theoretical framework in which a charged complex scalar field interacts with an electromagnetic gauge field, $A_{\mu}$, in a curved $(D+1)$-dimensional cosmic dispiration background. In a more general setting, a non-minimal coupling between the scalar field and the spacetime curvature may also be included. The quantum dynamics of the system is therefore governed by the following field equation:
\begin{eqnarray}
(\mathcal{D}^2 + m^2 + \xi R)\varphi(x)=0,
\label{E_motion}
\end{eqnarray}
where $m$ is the mass of the charged field $\varphi$, $R$ the Ricci scalar, $\xi$ an arbitrary numerical factor named curvature coupling and the diﬀerential operator $\mathcal{D}^2$ is defined as
\begin{eqnarray}
\mathcal{D}^2 = \frac{1}{\sqrt{|g|}}D_{\mu}\left(\sqrt{|g|}g^{\mu\nu}D_{\nu}\right),
\label{Diff_operator}
\end{eqnarray}
with $D_{\mu}=\partial_{\mu} + ieA_{\mu}$ and $g=\operatorname{det}\left(g_{\mu \nu}\right)$.

In the present analysis, the cosmic dispiration is treated as an idealized defect with a vanishingly small core. As a consequence, the scalar curvature $R$ is effectively described by a Dirac delta distribution, leading to a singular contribution localized at $r=0$, where the defect is situated~\cite{Vilenkin1994}. This feature would, in principle, allow for an interaction between the scalar field and the curvature through a non-minimal coupling term concentrated at the origin. To circumvent these localized contributions and keep the analysis as transparent as possible, we restrict ourselves to the minimally coupled case by setting $\xi = 0$. The role of non-minimal coupling in related contexts has been investigated in the literature; see, for instance, Ref.~\cite{Spinally:2000fjw}.

In order to proceed further, we consider a constant gauge field given by
\begin{eqnarray}
A_{\mu}=(0,0,A_{\phi},0),
\label{gauge_F}
\end{eqnarray}
where $A_{\phi}=-q\Phi_{\phi}/2\pi$, with $\Phi_{\phi}$ being a magnetic flux. The system under consideration, then, describes a magnetic flux threading the core of the cosmic dispiration, as schematically shown in Fig.~\ref{fig_dispiration}. As we will see, this configuration induces a nonvanishing azimuthal current density and, due to the helical structure of the spacetime, also generates a nonvanishing axial component, while the remaining components vanish.

By using the line element~\eqref{line_E1}, Eq.~\eqref{E_motion}, for $\xi=0$, can be expressed as
\begin{eqnarray}
\left[\frac{\partial^2}{\partial t^2}-\frac{1}{r} \frac{\partial}{\partial r}\left(r \frac{\partial}{\partial r}\right)-\frac{1}{r^2}\left(\frac{\partial}{\partial \phi}-\kappa \frac{\partial}{\partial z} + ieA_{\phi} \right)^2-\frac{\partial^2}{\partial z^2}-\sum_{i=4}^D \frac{\partial^2}{\partial x^{i 2}}+m^2\right] \varphi(x)=0.
\label{dif_eq}
\end{eqnarray}

Given that the Hamiltonian of the system commutes both with the angular momentum operator projected along the $z$-axis, also with $L_{\phi} = -i \partial_{\phi}$, and with the linear momentum operators $p_j = -i \partial_j$, for $j = 3, \ldots, D$, it is natural to exploit these symmetries by adopting the following {\it ansatz} in order to solve the partial differential equation~\eqref{dif_eq}, that is,
\begin{eqnarray}
\varphi_{\sigma}(x)=C R(r) e^{-i \omega_{\sigma} t+i n q \phi+i \nu z+i \mathbf{k} \cdot \mathbf{r}_{\|}},
\label{ansatz}
\end{eqnarray}
where $x=(t, r, \phi, z)$, $\sigma$ stands for the set of quantum numbers,  $C$ is a normalization constant and $\mathbf{r}_{\|}$and $\mathbf{k}$ represent, respectively, the coordinates of the extra dimensions and their corresponding momenta. Substituting Eq.~\eqref{ansatz} into~\eqref{dif_eq} we obtain the differential equation in $r$ as follows
\begin{equation}
\frac{d^2 R(r)}{dr^2} + \frac{1}{r}\frac{d R(r)}{dr}
+ \left(\eta^2 - \frac{\beta_{\sigma}^2}{r^2}\right) R(r) = 0,
\label{diff_r}
\end{equation}
where 
\begin{eqnarray}
\omega_{\sigma} &=& \sqrt{m^2 + \nu^2 + \eta^2 + {\bf k}^2},\nonumber\\
\beta_{\sigma}&=&q|(n - h + \alpha)|,\nonumber\\
\alpha&=&\frac{eA_{\phi}}{q},\nonumber\\
h&=&\frac{\kappa\nu}{q}.
\label{relations}
\end{eqnarray}
In this case, the set of quantum numbers is $\sigma=(\eta, n, \nu, {\bf k})$. Also, the parameter $\alpha$ above carries the magnetic flux information since it can also be written as $\alpha=-\frac{\Phi_{\phi}}{\Phi_0}$, with $\Phi_0=2\pi/e$ being the quantum flux. 

The differential equation in~\eqref{diff_r} is recognized as a Bessel-type equation, whose general solution can be expressed as a linear combination of the Bessel functions of the first kind, $J_{\mu}(x)$, and the second kind, $N_{\mu}(x)$. However, since the order $\beta_{\sigma}$ depends on the continuous quantum number $\nu$, the Neumann function typically exhibits a singular behavior at the origin, rendering it non-square-integrable. For this reason, physical considerations require us to discard this contribution and retain only the regular solution at the origin, given by the Bessel function $R(r)=J_{\beta_{\sigma}}(r)$, where $\beta_{\sigma}$ is defined in Eq.~\eqref{relations}.

Knowing the regular solution at origin, $r=0$, we can now determine the normalization constant $C$ in~\eqref{ansatz} by imposing the orthonormality condition
\begin{equation}
\int d^D x \, \sqrt{|g|} \, \varphi_{\sigma}(x)\varphi^*_{\sigma'}(x)
= \frac{1}{2\omega_{\sigma}}\,\delta_{\sigma,\sigma'} ,
\label{NC}
\end{equation}
where the symbol $\delta_{\sigma,\sigma'}$ denotes a Kronecker delta with respect to the discrete quantum number $n$, and Dirac delta distributions for the continuous parameters $\eta$, $\nu$, and $\mathbf{k}$. Substituting Eq.~\eqref{ansatz} into~\eqref{NC}, one finds that the normalization coefficient satisfies 
\begin{equation}
|C|^2 = \frac{q \eta}{2\omega_{\sigma} (2\pi)^{D-1}} .
\end{equation}

With this result, the properly normalized mode functions take the form
\begin{equation}
\varphi_{\sigma}(x) =
\left( \frac{q \eta}{2\omega_{\sigma} (2\pi)^{D-1}} \right)^{\frac{1}{2}}
J_{\beta_{\sigma}}(\eta r)\,
e^{-i\omega_\sigma t + i n q \phi + i \nu z + i \mathbf{k}\cdot \mathbf{r}_\parallel} .
\label{sol1}
\end{equation}
Alternatively, by expressing the solution in terms of the coordinate system $(t, r, \phi, Z, x_4, \ldots, x_D)$, one obtains a mode function that explicitly satisfies the identification condition given in Eq.~\eqref{IC}. In this representation, the general solution can be written as
\begin{equation}
\varphi_\sigma(x) =
\left( \frac{q \eta}{2\omega_\sigma (2\pi)^{D-1}} \right)^{\frac{1}{2}}
J_{\beta_\sigma}(\eta r)\,
e^{-i\omega_\sigma t + i(nq - \kappa \nu)\phi + i \nu Z + i \mathbf{k}\cdot \mathbf{r}_\parallel} .
\label{sol2}
\end{equation}
Note that the mode functions expressed in Eqs.~\eqref{sol1} and~\eqref{sol2} are physically equivalent, yielding identical observable quantities despite their different coordinate representations.

It is worth emphasizing that the structure of the mode functions derived above indicates that the quantity entering the order of the Bessel function in Eq.~\eqref{sol2} is effectively shifted according to
\begin{equation}
n \;\longrightarrow\; q(n - h + \alpha) ,
\end{equation}
when compared to the corresponding expression in Minkowski spacetime.
This modification admits a clear physical interpretation. The shift is not exclusively a consequence of the nontrivial topology of the background geometry, encoded in the parameters $\kappa$ and $q$, but also arises due to the presence of a magnetic flux, codified in $\alpha$. From a quantum-mechanical perspective, this can be interpreted as an effective redefinition of the angular momentum quantum number, analogous to the minimal coupling prescription in the presence of a vector potential, originally considered by Aharonov and Bohm~\cite{Aharonov:1959fk}.
Such a feature is a characteristic signature of an Aharonov--Bohm-type effect~\cite{Filgueiras:2005di, daSilva:2019lzh}, in which quantum modes acquire a phase that depends on the magnetic flux, even in regions where the associated field strength vanishes. Consequently, the system realizes a combined geometric and gauge-induced Aharonov--Bohm effect, where both the global properties of the spacetime and the magnetic flux contribute to observable quantum phenomena.

With the complete set of normalized mode functions at hand, we are now in a position to proceed further. In the following section, we will employ these solutions to construct the two-point Wightman function and subsequently evaluate the induced current density.
\section{Wightman function and Induced current density}
\label{secIV}
Following the standard field quantization procedure, the field operator can be written as
\begin{equation}
\hat{\varphi}(x) = \sum_\sigma\left(a_\sigma\varphi_\sigma(x) + a_\sigma^{\dagger}\varphi^*_\sigma(x)\right),
\label{F_quantization}
\end{equation}
where $\varphi_\sigma(x)$ is given by~\eqref{sol2}, and $\varphi^*_\sigma(x)$ denotes its complex conjugate. The operators $a_\sigma$ and $a_\sigma^{\dagger}$ are the annihilation and creation operators, respectively, satisfying the standard commutation relation
$[a_\sigma, a_{\sigma'}^{\dagger}] = \delta_{\sigma,\sigma'}$. The summation symbol above is defined as
\begin{equation}
\sum_\sigma \equiv \int d\mathbf{k}^{\, (D-3)} \int_{-\infty}^{\infty} d\nu \int_0^{\infty} d\eta \sum_{n=-\infty}^{\infty} .
\label{S_symbol}
\end{equation}
Once the vacuum state $|0\rangle$ is defined for the quantum field in~\eqref{F_quantization}, it follows that $a_\sigma|0\rangle = 0$ and $\langle 0|a_\sigma^{\dagger} = 0$.

The requirement that the radial solution given in~\eqref{sol2} remains regular at the origin, together with the interplay between the topological parameters $\kappa$ and $q$, and the magnetic flux encoded in $\alpha$, leads to a nontrivial modification of the vacuum structure. As a consequence, the spectrum of vacuum fluctuations, as well as the corresponding VEVs of physical observables, acquire a dependence on these parameters. To evaluate the impact of these modifications on the physical observables, it is essential to consider the positive-frequency Wightman function~\cite{BirrellDavies1982},
\begin{equation}
W(x,x') = \langle 0 | \hat{\varphi}(x)\hat{\varphi}^\dagger(x') | 0 \rangle,
\label{WF1}
\end{equation}
which plays a central role in our analysis. Expanding the field operator in terms of the complete set of normalized mode functions $\{\varphi_{\sigma}(x), \varphi_{\sigma}^*(x)\}$, as given in Eq.~\eqref{sol2}, the Wightman function can be expressed as the mode sum
\begin{equation}
W(x,x') = \sum_\sigma \varphi_\sigma(x)\varphi_\sigma^*(x') .
\label{WF2}
\end{equation}
Substituting Eq.~\eqref{sol2} into the expression above, the Wightman function can be written as
\begin{equation}
W(x,x')=
\frac{q}{2(2\pi)^{D-1}}
\sum_{\sigma}
\frac{e^{-i\omega_{\sigma}\Delta t+i {\bf k}\cdot \Delta {\bf r}_{\parallel}}}{\omega_{\sigma}}
e^{iq(n-h)\Delta\phi+i\nu\Delta Z}
\eta
J_{\beta_{\sigma}}(\eta r)
J_{\beta_{\sigma}}(\eta r') ,
\label{WF3}
\end{equation}
where $\Delta t=t-t'$, $\Delta Z=Z-Z'$, $\Delta {\bf r}_{\parallel}=r_{\parallel}-r'_{\parallel}$ and $\Delta\phi=\phi - \phi'$.
To proceed with the evaluation of the expression above, we make use of the following integral representation
\begin{equation}
\frac{e^{-\omega_{\sigma}\Delta\tau}}{\omega_{\sigma}}
=
\frac{2}{\sqrt{\pi}}
\int_{0}^{\infty}
ds\,
e^{-s^2\omega_{\sigma}^2-\frac{\Delta\tau^2}{4s^2}} .
\label{Int_re}
\end{equation}
By setting $\Delta\tau=i\Delta t$ (Wick rotation) and substituting this representation into Eq.~\eqref{WF3}, and taking into account~\eqref{S_symbol}, the integration over the continuous modes ${\bf k}=(k_4,\ldots,k_D)$ can be performed. The integration over the variable $\eta$ is then carried out by employing the identity
\begin{equation}
\int_{0}^{\infty} d\eta \, \eta
J_{\gamma}(\eta r)
J_{\gamma}(\eta r')
e^{-s^2\eta^2}
=
\frac{1}{2s^2}
e^{-\frac{r^2+r'^2}{4s^2}}
I_{\gamma}(w),
\label{identity1}
\end{equation}
where $w=\frac{rr'}{2s^2}$ and $I_{\gamma}(x)$ is the modified Bessel function of the first kind. After considering these steps, the Wightman function can be expressed as
\begin{equation}
W(x,x')=
\frac{q}{2(2\pi)^{\frac{D+2}{2}}(rr')^{\frac{D-2}{2}}}
\int_{0}^{\infty}
dw\, w^{\frac{D-4}{2}}
e^{-\frac{m^2rr'}{2w}-\frac{w\Delta\zeta^2}{2rr'}}
\mathcal{I}(w,\kappa,q),
\label{WF4}
\end{equation}
where $\Delta\zeta^2=\Delta\tau^2 + \Delta {\bf r}^2_{||} + r^2 + r'^2$ and the function $\mathcal{I}(w,\kappa,q)$ is defined by
\begin{equation}
\mathcal{I}(w,\kappa,q)=\int_{-\infty}^{\infty}\frac{q\,dh}{\kappa}
e^{-\frac{q^2rr' h^2}{2w\kappa^2} + i\frac{qh\Delta Z}{\kappa}}
\sum_{n=-\infty}^{\infty}
e^{iq(n-h)\Delta\phi}
I_{\beta_{\sigma}}(w).
\label{I_fun}
\end{equation}
The representation of the Wightman function given in Eq.~\eqref{WF4} will be useful for the evaluation of the induced current density in the next section.

We now focus on the evaluation of the VEV of the bosonic current density operator, given by
\begin{eqnarray}
\begin{aligned}
j_\mu(x) & = i e\left[\hat{\varphi}^*(x) D_\mu \hat{\varphi}(x)-\left(D_\mu \hat{\varphi}\right)^* \hat{\varphi}(x)\right] \\
& = i e\left[\hat{\varphi}^*(x) \partial_\mu \hat{\varphi}(x)-\hat{\varphi}(x)\left(\partial_\mu \hat{\varphi}(x)\right)^*\right]-2 e^2 A_\mu(x)|\hat{\varphi}(x)|^2.
\end{aligned}
\label{BC_def}
\end{eqnarray}
Its VEV can be evaluated in terms of the positive-frequency Wightman function as
\begin{eqnarray}
\left\langle j_\mu(x)\right\rangle = i e \lim_{x^{\prime} \rightarrow x}\left\{\left(\partial_\mu-\partial_\mu^{\prime}\right) W\left(x, x^{\prime}\right)+2 i e A_\mu(x) W\left(x, x^{\prime}\right)\right\}.
\label{BC_vev}
\end{eqnarray}
This expression is a periodic function of the magnetic flux $\Phi_\phi$, with period equal to the quantum flux. This property becomes explicit by expressing the parameter $\alpha$ in~\eqref{relations} as
\begin{eqnarray}
\alpha = n_0 + \alpha_0, \quad \text{with} \quad |\alpha_0| < \frac{1}{2},
\label{fractional}
\end{eqnarray}
where $n_0$ is an integer that can be absorbed into a redefinition of the summation index $n$ appearing in~\eqref{I_fun}. With this decomposition, the VEV of the current density depends only on the fractional part $\alpha_0$, being insensitive to the integer contribution. As a consequence, the dependence on $\alpha_0$ is periodic, reflecting the underlying Aharonov--Bohm nature of the effect, as already discussed.

The components of the induced current~\eqref{BC_vev} all vanish, except for the azimuthal and axial components. This behavior is consistent with the symmetry of the physical system, which consists of a cosmic dispiration spacetime threaded by a magnetic flux along the core of the defect, as illustrated in Fig.~\ref{fig_dispiration}. While the azimuthal component is associated with the circular symmetry around the defect, the axial component arises due to the helical structure of the spacetime. 

Before proceeding with the evaluation of the vacuum current density, let us briefly comment on its ultraviolet behavior. Although the Wightman function in Eq.~\eqref{WF4} contains the standard local singularity, the corresponding Minkowski contribution to the current density vanishes identically. Indeed, after the action of the operator $(\partial_\mu-\partial'_\mu)$ and taking the coincidence limit, the Minkowski contribution has the generic structure

\begin{equation}
\langle j_\mu \rangle_M \propto \int d^Dk \, k_\mu \, F(k^2),
\end{equation}
where $F(k^2)$ is an even function of the momentum. Since the integrand is odd under $k_\mu \rightarrow -k_\mu$, the integral vanishes by parity. Hence, the local Minkowski sector does not contribute to the vacuum current density, and the resulting current is entirely determined by the finite topological contribution induced by the cosmic dispiration and the magnetic flux. This contrasts with observables such as the vacuum energy density, for which the corresponding momentum integrals contain even powers of the momentum and generally require ultraviolet renormalization.

\subsection{Azimuthal current density}
A nonvanishing contribution arises from the azimuthal component, $\left\langle j_\phi(x)\right\rangle$, which can be obtained by rewriting Eq.~\eqref{BC_vev} as
\begin{eqnarray}
\left\langle j_\phi(x)\right\rangle = 2 i e \lim_{x^{\prime} \rightarrow x}\left\{\partial_{\Delta\phi} W\left(x, x^{\prime}\right)+ i e A_\phi(x) W\left(x, x^{\prime}\right)\right\}.
\label{BC_def_azimuthal}
\end{eqnarray}
Differentiating Eq.~\eqref{WF4} with respect to $\Delta\phi$ and taking the coincidence limit, $x' \to x$, we obtain
\begin{equation}
\left\langle j_\phi(x)\right\rangle =
-
\frac{e q}{(2\pi)^{\frac{D+2}{2}} r^{D-2}}
\int_{0}^{\infty}
dw\, w^{\frac{D-4}{2}}
e^{-\frac{m^2 r^2}{2w}-w}\mathcal{J}_{\phi}(w),
\label{BC_def_azimuthal2}
\end{equation}
where
\begin{equation}
\mathcal{J}_{\phi}(w)=\int_{-\infty}^{\infty}\frac{q\,dh}{\kappa}
e^{-\frac{q^2 r^2 h^2}{2w\kappa^2}}
\sum_{n=-\infty}^{\infty}\beta_{\sigma}
I_{\beta_{\sigma}}(w).
\label{J_fun}
\end{equation}

In Appendix~\ref{Ap_a}, we derive a more convenient expression for Eq.~\eqref{J_fun}, given by Eq.~\eqref{J_fun_text}. After substituting Eq.~\eqref{J_fun_text} into Eq.~\eqref{BC_def_azimuthal2}, we will have to perform the integral in $w$. This can be done with the aid of the relation
\begin{equation}
\int_{0}^{\infty}
dw w^{{\frac{D-1}{2}}}
e^{-\frac{m^2r^2}{2w}-\frac{p^2j^2w}{2r^2} - 2ws^2}=2(mr)^{D+1}f_{\frac{D+1}{2}}\left(m\sqrt{(jp)^2 + (2rs)^2}\right),
\label{int1_A}
\end{equation}
where we have defined
\begin{equation}
f_{b}(x) = \frac{K_{b}(x)}{x^{b}},
\label{def0_A}
\end{equation}
with $K_{b}(x)$ being the Macdonald function. Hence, the azimuthal contravariant current density is obtained as
\begin{eqnarray}
\left\langle j^\phi(x)\right\rangle&=&
\frac{4em^{D+1}}{(2\pi)^{\frac{D+1}{2}}}\left\{\sum_{\ell=1}^{[q/2]}\sin(2\ell\pi/q)\sin(2\ell\pi\alpha_0)f_{\frac{D+1}{2}}\left(m\sqrt{(p\ell)^2 + (2rs_{\ell})^2}\right)\right.\nonumber\\
&+&\left.\frac{q}{2\pi^2}\int_0^{\infty}dy\sinh y\sum_{n=-\infty}^{\infty}S_n(y,\alpha_0,q)f_{\frac{D+1}{2}}\left(m\sqrt{(pn)^2 + (2rs_{y})^2}\right)\right\},
\label{BC_def_azimuthal_final}
\end{eqnarray}
where $s_{\ell}=\sin(\ell\pi/q)$, $s_y=\cosh(y/2)$, and $[q/2]$ denotes the integer part of $q/2$. In the special case $q = 2g$, with $g$ being an integer, the term corresponding to $\ell = q/2$ must be taken with a weight factor of $1/2$. The function $S_n$ is given by
\begin{eqnarray}
S_n(y,\alpha_0,q)=\frac{\left(\frac{qy}{2\pi}\right)\sin(2\pi n\alpha_0)}{\left(n - \frac{q}{2}\right)^2 + \left(\frac{qy}{2\pi}\right)^2}.
\label{sum1}
\end{eqnarray}
For large values of $mr \gg 1$ and $m\kappa \gg 1$, or when one of these parameters is kept fixed while the other increases, the induced current~\eqref{BC_def_azimuthal_final} is exponentially suppressed due to the asymptotic behavior of the Macdonald function for large arguments, namely,
$K_{b}(x)\simeq\sqrt{\frac{\pi}{2x}}e^{-x}$~\cite{abramowitz1965handbook}. This behavior is illustrated, for $D=3$, in the bottom panels of Fig.~\ref{fig2}. In contrast, the small-argument limit yields the massless case for the induced current density, as will be discussed next.

The massless limit of the current density in Eq.~\eqref{BC_def_azimuthal_final} can be obtained by making use of the following relation~\cite{abramowitz1965handbook}:
\begin{eqnarray}
\lim_{z \to 0} z^{b} K_b(nz) &=& \frac{2^{b -1}\Gamma(b)}{n^{b}}\nonumber\\
&=&\frac{2^{\frac{D-1}{2}}\Gamma\left(\frac{D+1}{2}\right)}{n^{\frac{D+1}{2}}},
\label{lim}
\end{eqnarray}
where $b=\frac{D+1}{2}$ in the present case. This leads to
\begin{eqnarray}
\left\langle j^\phi(x)\right\rangle&=&
\frac{2e\Gamma\left(\frac{D+1}{2}\right)}{\pi^{\frac{D+1}{2}}}\left\{\sum_{\ell=1}^{[q/2]}\frac{\sin(2\pi \ell/q)\sin(2\pi \ell\alpha_0)}{\left[(p\ell)^2 + (2rs_\ell)^2\right]^{\frac{D+1}{2}}}\right.\nonumber\\
&+&\left.\frac{q}{2\pi^2}\int_0^{\infty}dy\,\sinh y\sum_{n=-\infty}^{\infty}\frac{S_n(y,\alpha_0,q)}{\left[(pn)^2 + (2rs_y)^2\right]^{\frac{D+1}{2}}}\right\}.
\label{BC_def_azimuthal_final_massless}
\end{eqnarray}
Thus, Eqs.~\eqref{BC_def_azimuthal_final} and~\eqref{BC_def_azimuthal_final_massless} describe the induced azimuthal currents for the massive and massless charged scalar fields, respectively. As can be seen, both expressions encode the effects of the spacetime topology and the magnetic flux through the parameters $q$, $\kappa$, and $\alpha_0$. In particular, for $\alpha_0=0$ and $\alpha_0=1/2$, the induced currents vanish. Also in both expressions, for $q<2$, the first term on the r.h.s. is absent, and only the second term, involving the integral over $y$, contributes. Specifically, for $q=1$, only the contribution due to the screw dislocation survives through the second term. Another noteworthy feature is that, for $\kappa \neq 0$, the induced currents remain finite and nonzero at the origin, $r=0$. In the massless case, Eq.~\eqref{BC_def_azimuthal_final_massless} behaves as $\left\langle j^\phi(x)\right\rangle \propto \kappa^{-(D+1)}$.

The case $\kappa = 0$ corresponds to the induced current in a cosmic string spacetime in the presence of a magnetic flux and has been previously investigated in Refs.~\cite{Braganca:2014qma, BezerradeMello:2014phm}. In Appendix~\ref{Ap_series}, we demonstrate that the sum over $n$ of the function $S_n(y,\alpha_0,q)$ in Eqs.~\eqref{BC_def_azimuthal_final} and~\eqref{BC_def_azimuthal_final_massless}, upon setting $\kappa = 0$, can be expressed in closed form in terms of hyperbolic functions, as given in Eq.~\eqref{B21}. Substituting this result into Eqs.~\eqref{BC_def_azimuthal_final} and~\eqref{BC_def_azimuthal_final_massless}, we recover the expressions obtained in Refs.~\cite{Braganca:2014qma, BezerradeMello:2014phm}. Therefore, our results are in full agreement with those previously reported in the literature.
\subsubsection{Case $D=3$}
\begin{figure}
    \centering
    \includegraphics[width=0.4\textwidth]{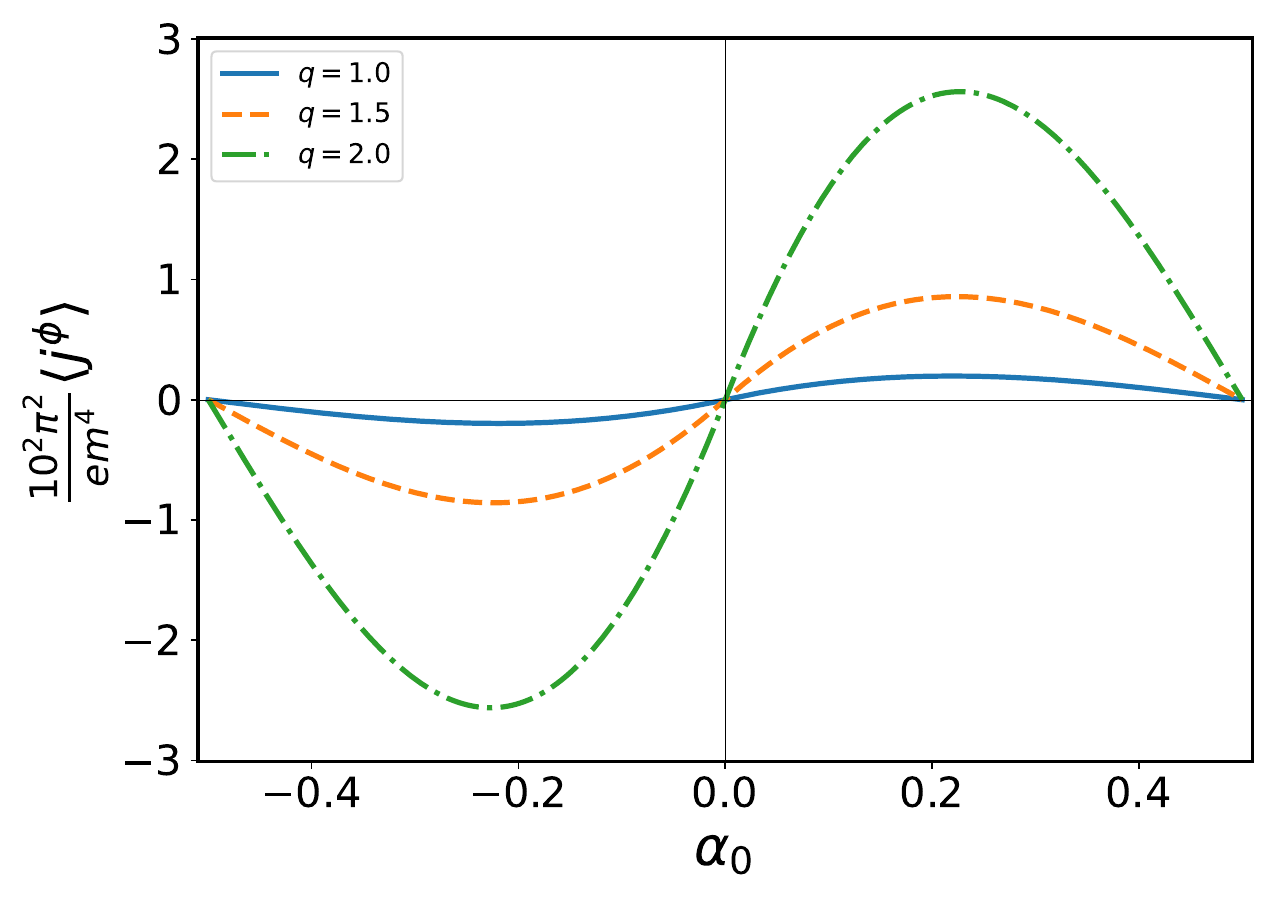}
    \includegraphics[width=0.4\textwidth]{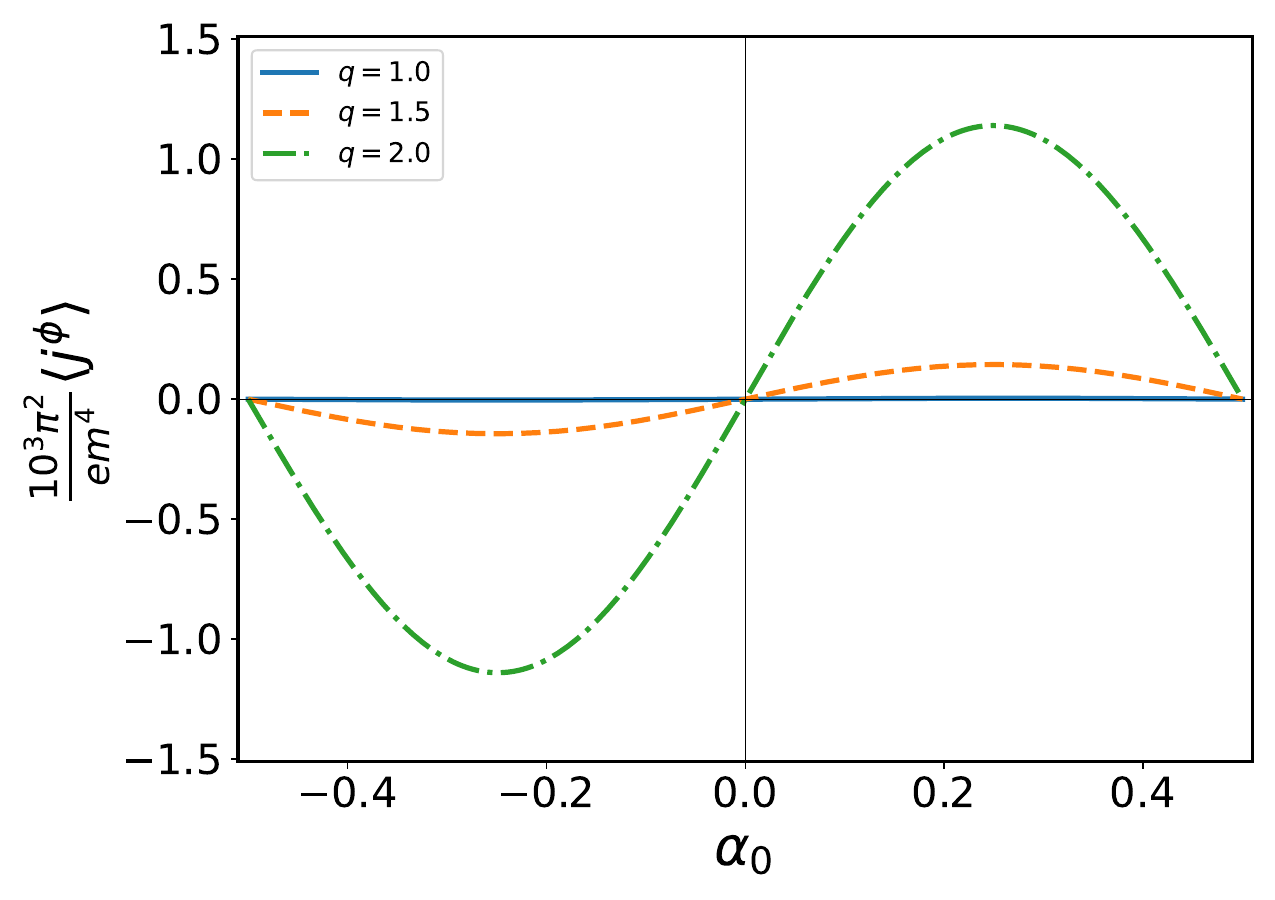}
    \includegraphics[width=0.4\textwidth]{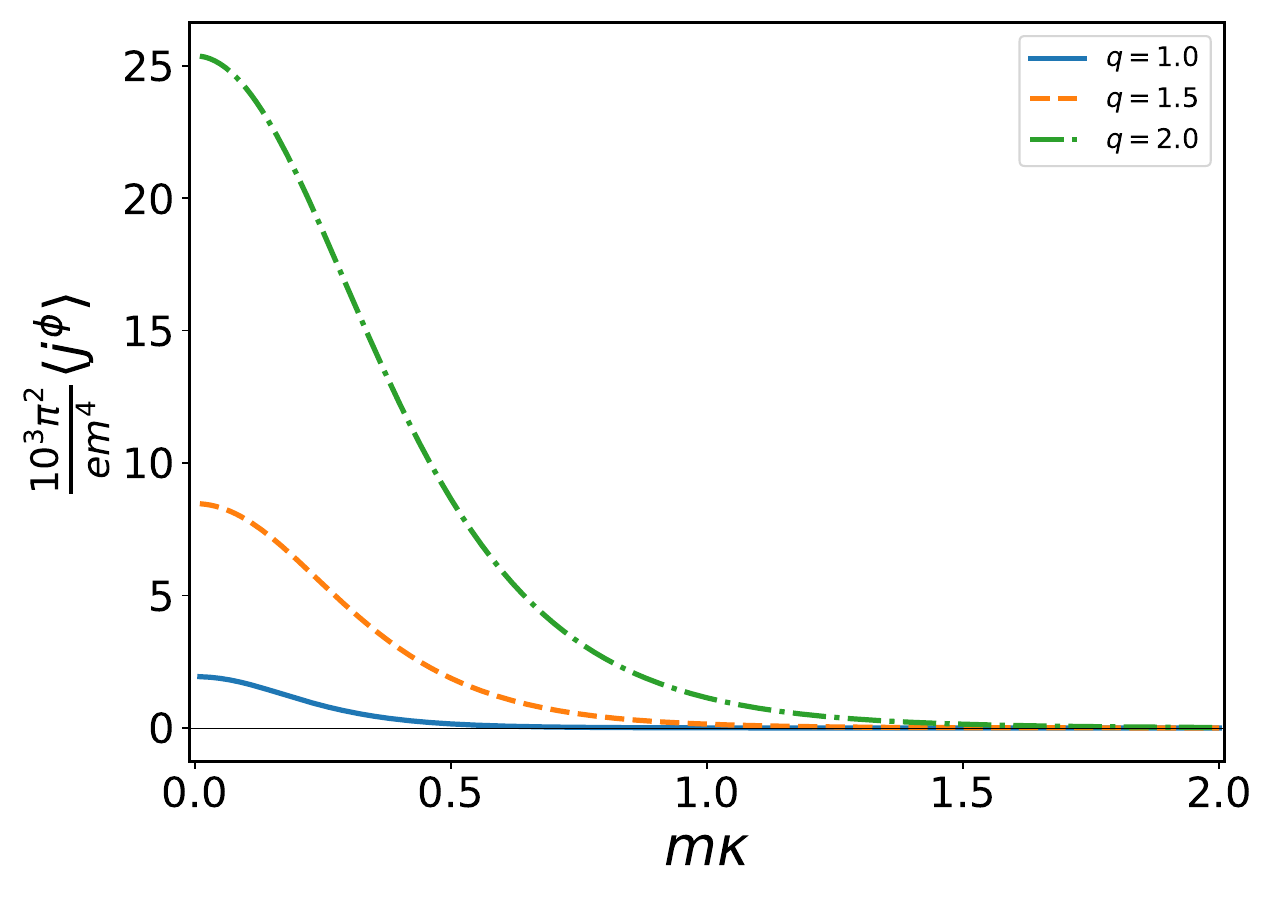}
    \includegraphics[width=0.4\textwidth]{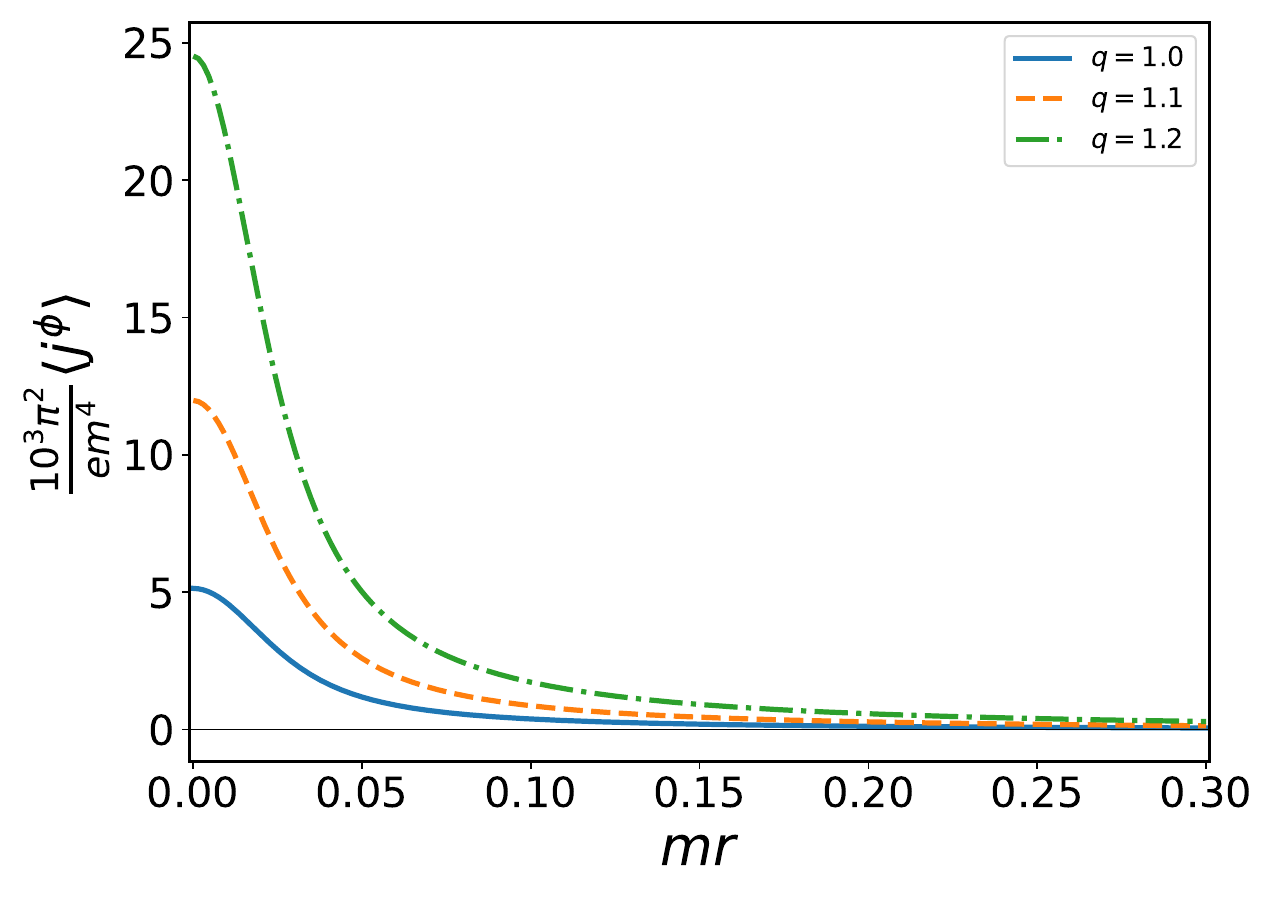}
    \caption{Plot of the induced current~\eqref{BC_def_azimuthal_final_D=3} in dimensionless form as a function of $\alpha_0$ (top panels) and as a function of $m\kappa$ and $mr$ (bottom panels), for different values of $q$. In the top left panel, we set $mr=1$ and $m\kappa=0$, whereas in the top right panel we take $mr=m\kappa=1$. In the bottom panels, we fix $\alpha_0=0.25$, with $mr=1$ in the left panel and $m\kappa=1$ in the right panel.}
    \label{fig2}
\end{figure}

%
We now turn to the analysis of the induced current in the $(3+1)$-dimensional spacetime. By setting $D=3$ in Eq.~\eqref{BC_def_azimuthal_final}, we obtain the corresponding expression for the azimuthal current associated with a massive scalar field. This particular case allows for a more direct inspection of the induced current in a physically relevant system. In this case, we find
\begin{eqnarray}
\left\langle j^\phi(x)\right\rangle&=&
\frac{e m^{4}}{\pi^{2}}\left\{\sum_{\ell=1}^{[q/2]}\sin(2\pi\ell/q)\sin(2\pi \ell\alpha_0)f_{2}\left(m\sqrt{(p\ell)^2 + (2rs_{\ell})^2}\right)\right.\nonumber\\
&+&\left.\frac{q}{2\pi^2}\int_0^{\infty}dy\,\sinh y\sum_{n=-\infty}^{\infty}S_n(y,\alpha_0,q)f_{2}\left(m\sqrt{(pn)^2 + (2rs_{y})^2}\right)\right\}.
\label{BC_def_azimuthal_final_D=3}
\end{eqnarray}
This expression is displayed in Fig.~\ref{fig2} in dimensionless form as a function of $\alpha_0$, $m\kappa$, and $mr$, for different values of $q$, as indicated in the legend. The top panels show the periodic dependence of the induced current on $\alpha_0$, as expected. In the top left panel, we set $mr=1$ and $m\kappa=0$, whereas in the top right panel we take $mr=m\kappa=1$. A comparison between these panels shows that the presence of the screw dislocation parameter $\kappa$ leads to a suppression of the magnitude of the induced current.

In the bottom panels, where the induced current is shown as a function of $m\kappa$ and $mr$, we fix $mr=1$ in the left panel and $m\kappa=1$ in the right panel. In both cases, we take $\alpha_0=0.25$. These plots indicate the suppression of the induced current for large values of $m\kappa$ and $mr$. At $m\kappa=0$ (or $mr=0$), corresponding to the horizontal axis, the current approaches constant values associated with the massless limit.

It is worth noting that a divergence arises when both $\kappa$ and $r$ vanish in Eq.~\eqref{BC_def_azimuthal_final_D=3}. Accordingly, in the bottom left panel, if we considered smaller values of $mr$, the curves would tend to diverge as $m\kappa \to 0$, whereas for larger values of $mr$ the suppression would become more pronounced. In contrast, in the bottom right panel, if we took small values of $m\kappa$, the curves would tend to diverge as $mr \to 0$, while for larger values of $m\kappa$ the suppression would become enhanced.

For $D=3$, the massless scalar field case follows from Eq.~\eqref{BC_def_azimuthal_final_massless} and can be written as
\begin{eqnarray}
\left\langle j^\phi(x)\right\rangle&=&
\frac{2e}{\pi^{2}}\left\{\sum_{\ell=1}^{[q/2]}\frac{\sin(2\ell\pi/q)\sin(2\ell\pi\alpha_0)}{\left[(p\ell)^2 + (2rs_\ell)^2\right]^{2}}\right.\nonumber\\
&+&\left.\frac{q}{2\pi^2}\int_0^{\infty}dy\sinh y\sum_{n=-\infty}^{\infty}\frac{S_n(y,\alpha_0,q)}{\left[(pn)^2 + (2rs_y)^2\right]^{2}}\right\}.
\label{BC_def_azimuthal_final_massless_D=3}
\end{eqnarray}
\begin{figure}
    \centering
    \includegraphics[width=0.4\textwidth]{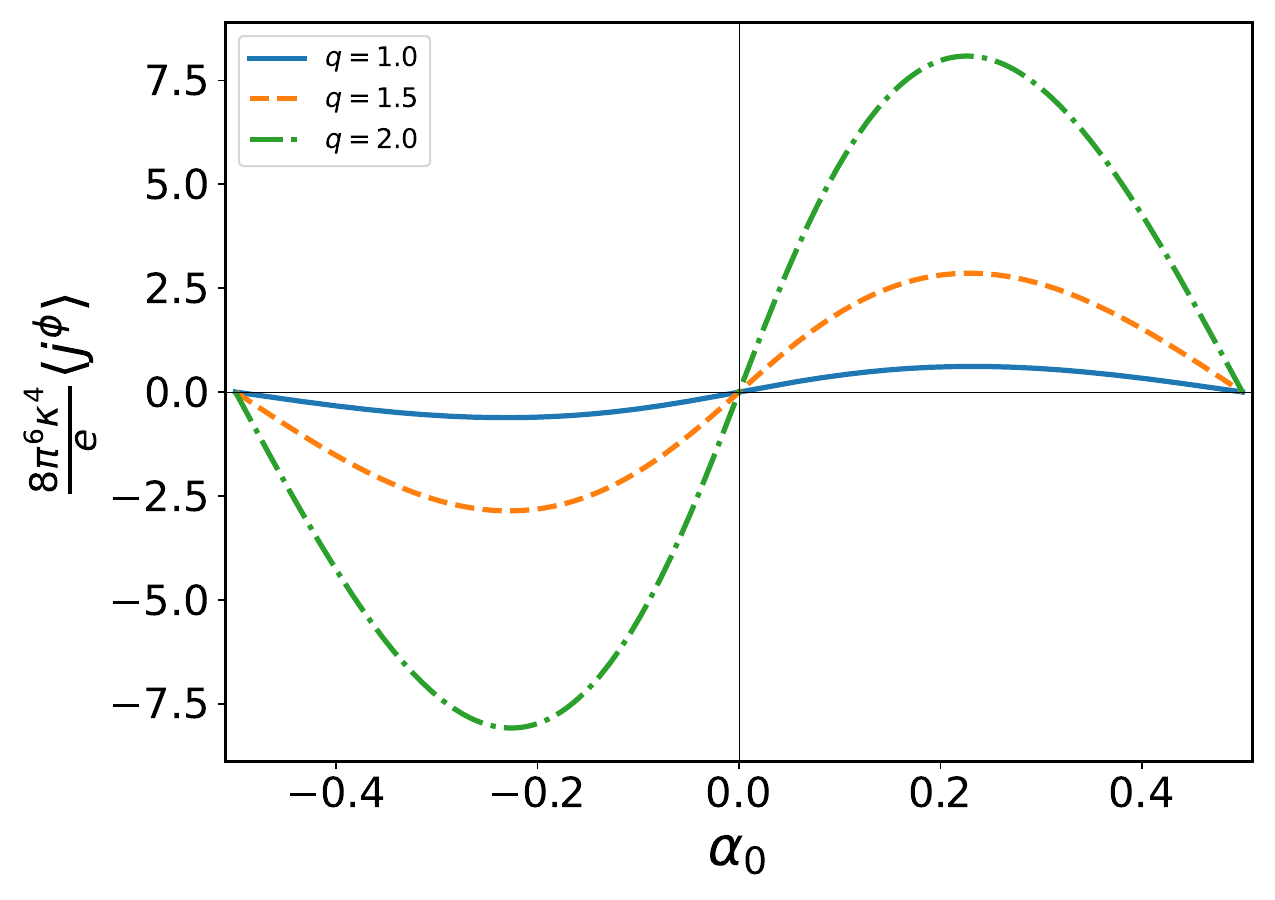}
    \includegraphics[width=0.4\textwidth]{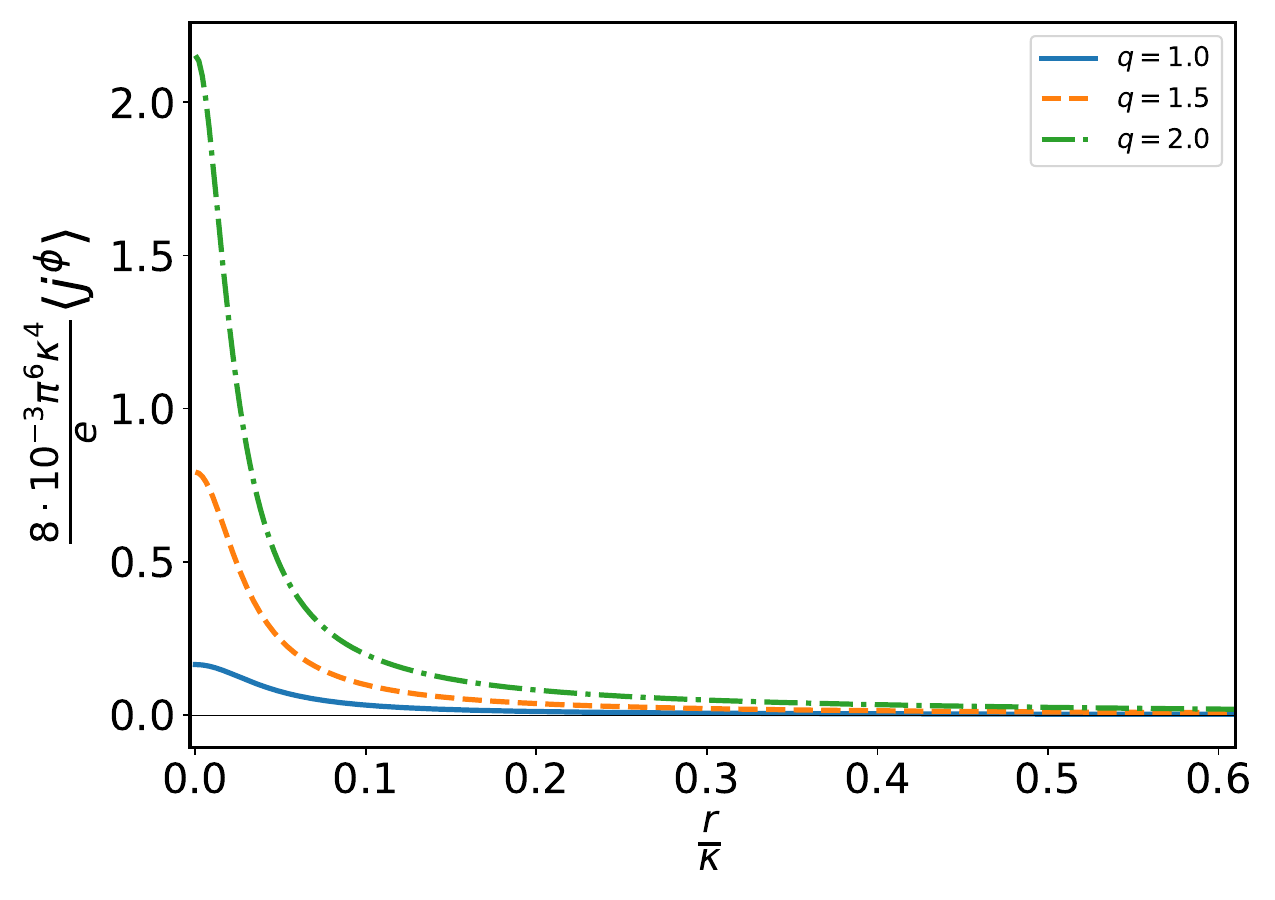}
    \caption{Plot of the induced current~\eqref{BC_def_azimuthal_final_massless_D=3} in dimensionless form as a function of $\alpha_0$ (left panel) and of $r/\kappa$ (right panel), for different values of $q$. In the left panel, we fix $r/\kappa=1$, whereas in the right panel we take $\alpha_0=0.25$.}
 \label{fig3}
\end{figure}

This expression is displayed in Fig.~\ref{fig3} in dimensionless form as a function of $\alpha_0$ (left panel) and of $r/\kappa$ (right panel). In the left panel, we fix $r/\kappa=1$, whereas in the right panel we take $\alpha_0=0.25$. The periodic dependence of the induced current on $\alpha_0$ is clearly observed, along with a strong suppression for large values of $r/\kappa$. At $r/\kappa=0$, the current attains values consistent with the behavior $\left\langle j^\phi(x)\right\rangle \propto \kappa^{-4}$.

\subsection{Axial current density}
As a consequence of the helical structure of the defect, an additional nonvanishing contribution emerges from the axial component of the current, $\left\langle j_{Z}(x)\right\rangle$. This component can be obtained by rewriting Eq.~\eqref{BC_vev} in the form
\begin{eqnarray}
\left\langle j_{Z}(x)\right\rangle = 2 i e \lim_{x^{\prime} \rightarrow x}\partial_{\Delta Z} W\left(x, x^{\prime}\right).
\label{BC_def_Axial}
\end{eqnarray}
By differentiating Eq.~\eqref{WF4} with respect to $\Delta Z$ and subsequently taking the coincidence limit, $x' \to x$, we obtain
\begin{equation}
\left\langle j_{Z}(x)\right\rangle =
-
\frac{e q}{(2\pi)^{\frac{D+2}{2}} r^{D-2}}
\int_{0}^{\infty}
dw\, w^{\frac{D-4}{2}}
e^{-\frac{m^2 r^2}{2w}-w}\mathcal{J}_Z(w),
\label{BC_def_Axial2}
\end{equation}
where
\begin{equation}
\mathcal{J}_Z(w)=\frac{q^2}{\kappa^2}\int_{-\infty}^{\infty}hdh
e^{-\frac{q^2 r^2 h^2}{2w\kappa^2}}
\sum_{n=-\infty}^{\infty}
I_{\beta_{\sigma}}(w).
\label{J_fun_Z}
\end{equation}

In Appendix~\ref{Ap_axial}, we also derive a more convenient representation for $\mathcal{J}_Z(w)$, given in Eq.~\eqref{J_axial2}. Substituting this result into Eq.~\eqref{BC_def_Axial2}, the remaining integral over $w$ can be evaluated with the aid of the relation~\eqref{int1_A}. In this way, for the contravariant component, we find
\begin{eqnarray}
\left\langle j^Z(x)\right\rangle&=&
\frac{4\kappa em^{D+1}}{q(2\pi)^{\frac{D-1}{2}}}\left\{\sum_{\ell=1}^{[q/2]}\ell\sin(2\ell\pi\alpha_0)f_{\frac{D+1}{2}}\left(m\sqrt{(p\ell)^2 + (2rs_{\ell})^2}\right)\right.\nonumber\\
&+&\left.\frac{q}{2\pi^2}\int_0^{\infty}dy\sum_{n=-\infty}^{\infty}M_n(y,\alpha_0,q)f_{\frac{D+1}{2}}\left(m\sqrt{(pn)^2 + (2rs_{y})^2}\right)\right\},
\label{BC_def_Axial2_final}
\end{eqnarray}
where
\begin{eqnarray}
M_n(y,\alpha_0,q)=\frac{\left(n - \frac{q}{2}\right)n\sin(2\pi n\alpha_0)}{\left(n - \frac{q}{2}\right)^2 + \left(\frac{qy}{2\pi}\right)^2}.
\label{sum_M1}
\end{eqnarray}
The sum over $\ell$ above obeys the same restrictions as the corresponding sum appearing in Eq.~\eqref{BC_def_azimuthal_final}. Moreover, for large values of $mr \gg 1$ and $m\kappa \gg 1$, or when one of these parameters is kept fixed while the other increases, the induced current~\eqref{BC_def_Axial2_final} is exponentially suppressed. This behavior follows from the asymptotic form of the Macdonald function for large arguments~\cite{abramowitz1965handbook}, as already discussed for the azimuthal component. 

The massless limit of the axial current density in Eq.~\eqref{BC_def_Axial2_final} can be obtained by making use of the limiting expression given in Eq.~\eqref{lim}. Consequently, we obtain
\begin{eqnarray}
\left\langle j^Z(x)\right\rangle&=&
\frac{4\kappa e\Gamma\left(\frac{D+1}{2}\right)}{q\pi^{\frac{D-1}{2}}}\left\{\sum_{\ell=1}^{[q/2]}\frac{\ell\sin(2\pi \ell\alpha_0)}{\left[(p\ell)^2 + (2rs_\ell)^2\right]^{\frac{D+1}{2}}}\right.\nonumber\\
&+&\left.\frac{q}{2\pi^2}\int_0^{\infty}dy\sum_{n=-\infty}^{\infty}\frac{M_n(y,\alpha_0,q)}{\left[(pn)^2 + (2rs_y)^2\right]^{\frac{D+1}{2}}}\right\}.
\label{BC_def_axial_final_massless}
\end{eqnarray}
It is worth noting that, upon setting $\kappa=0$, the axial current densities~\eqref{BC_def_Axial2_final} and~\eqref{BC_def_axial_final_massless} vanish identically. This result reflects the decoupling between the $\phi$ and $Z$ coordinates, in which case the only nonvanishing contributions correspond to the azimuthal components given in Eqs.~\eqref{BC_def_azimuthal_final} and~\eqref{BC_def_azimuthal_final_massless} for $\kappa=0$. 
Therefore, Eqs.~\eqref{BC_def_Axial2_final} and~\eqref{BC_def_axial_final_massless} describe the induced axial currents for massive and massless charged scalar fields, respectively. These expressions, similarly to the azimuthal component, encode the effects of both the nontrivial spacetime topology and the magnetic flux through the parameters $q$, $\kappa$, and $\alpha_0$. In particular, the axial currents also vanish for $\alpha_0=0$ and $\alpha_0=1/2$.
For $\kappa \neq 0$, the induced axial currents remain finite and nonvanishing at the origin, $r=0$. In the massless case, Eq.~\eqref{BC_def_axial_final_massless} exhibits the behavior $\left\langle j^Z(x)\right\rangle \propto \kappa^{-D}$.
\begin{figure}
    \centering
    \includegraphics[width=0.4\textwidth]{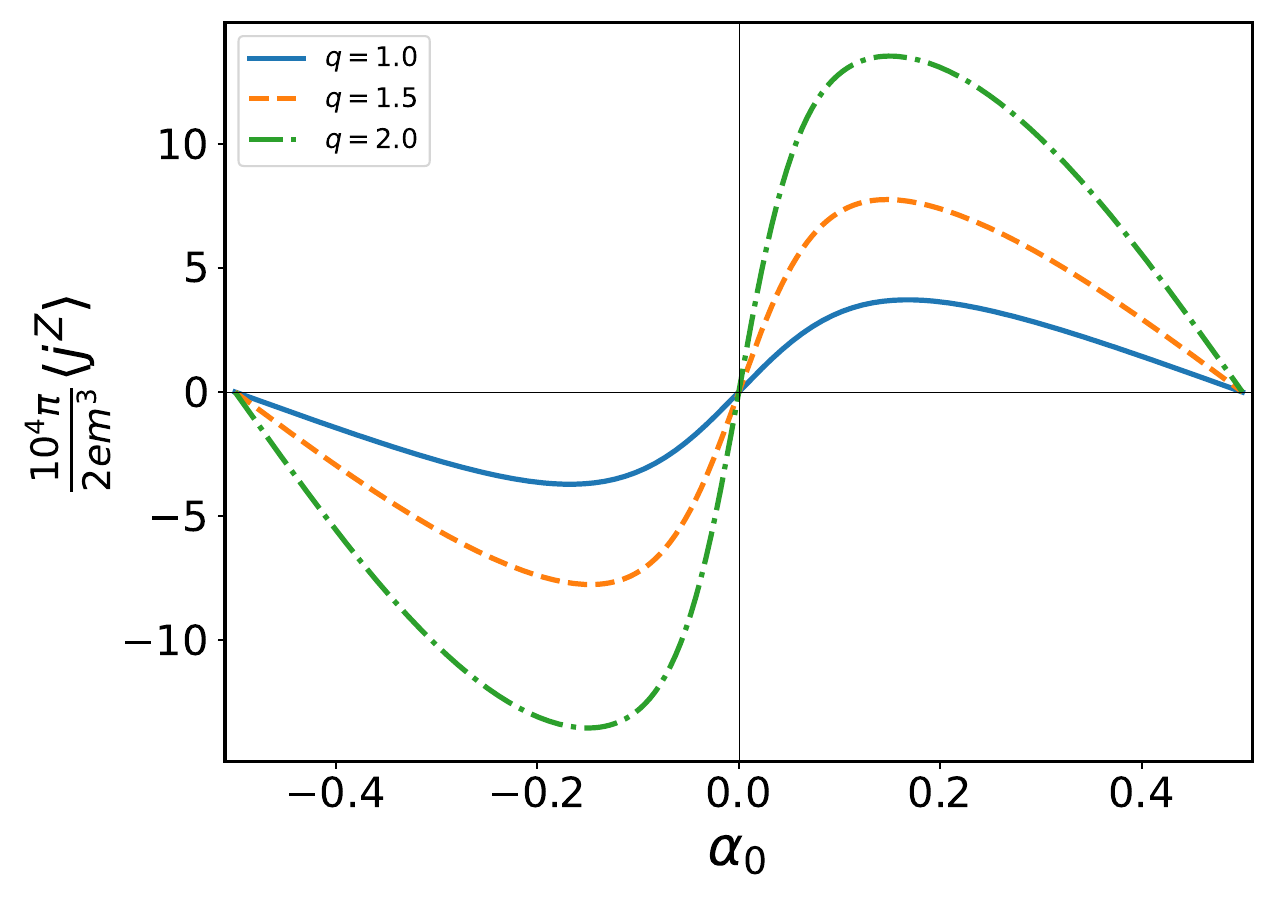}
    \includegraphics[width=0.4\textwidth]{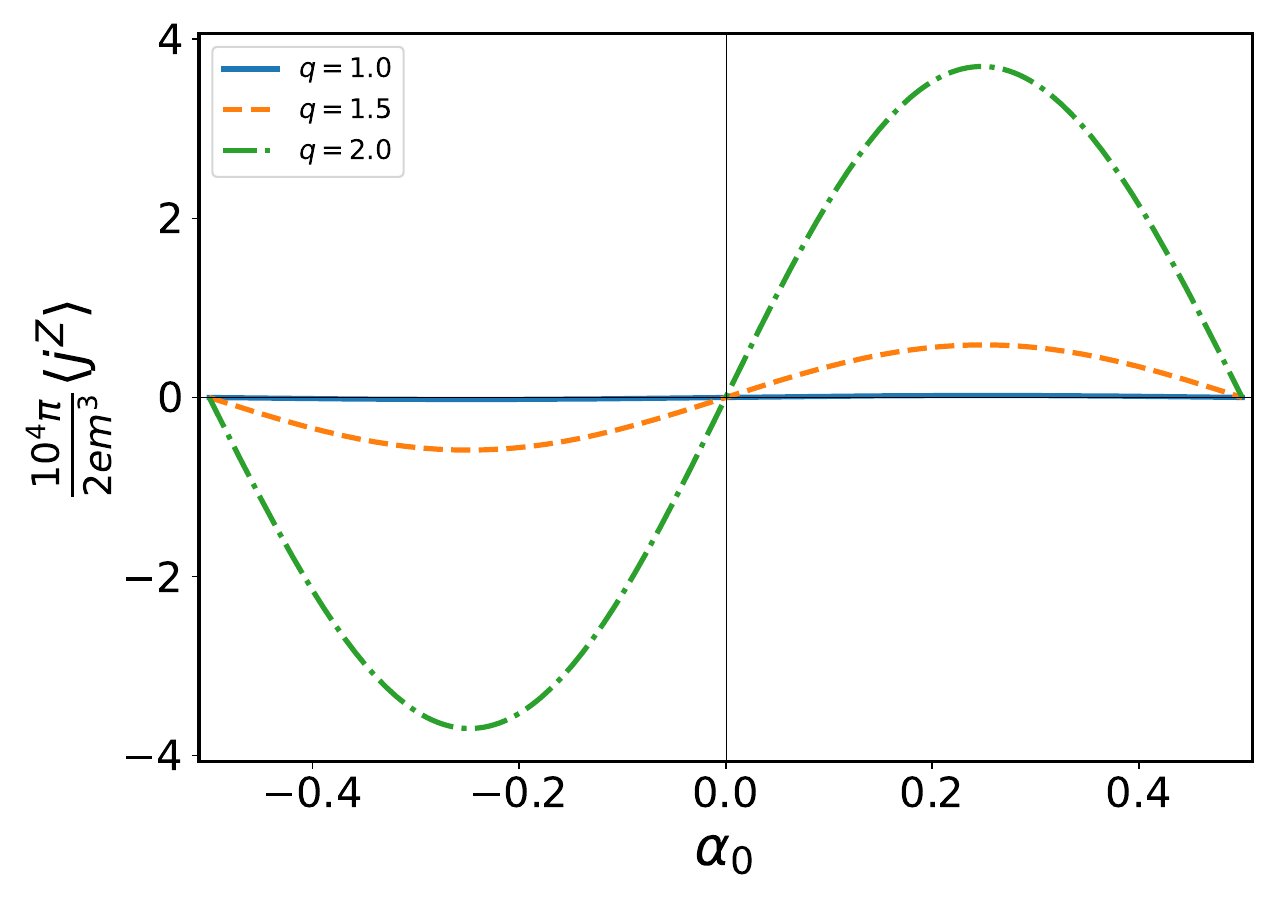}
    \includegraphics[width=0.4\textwidth]{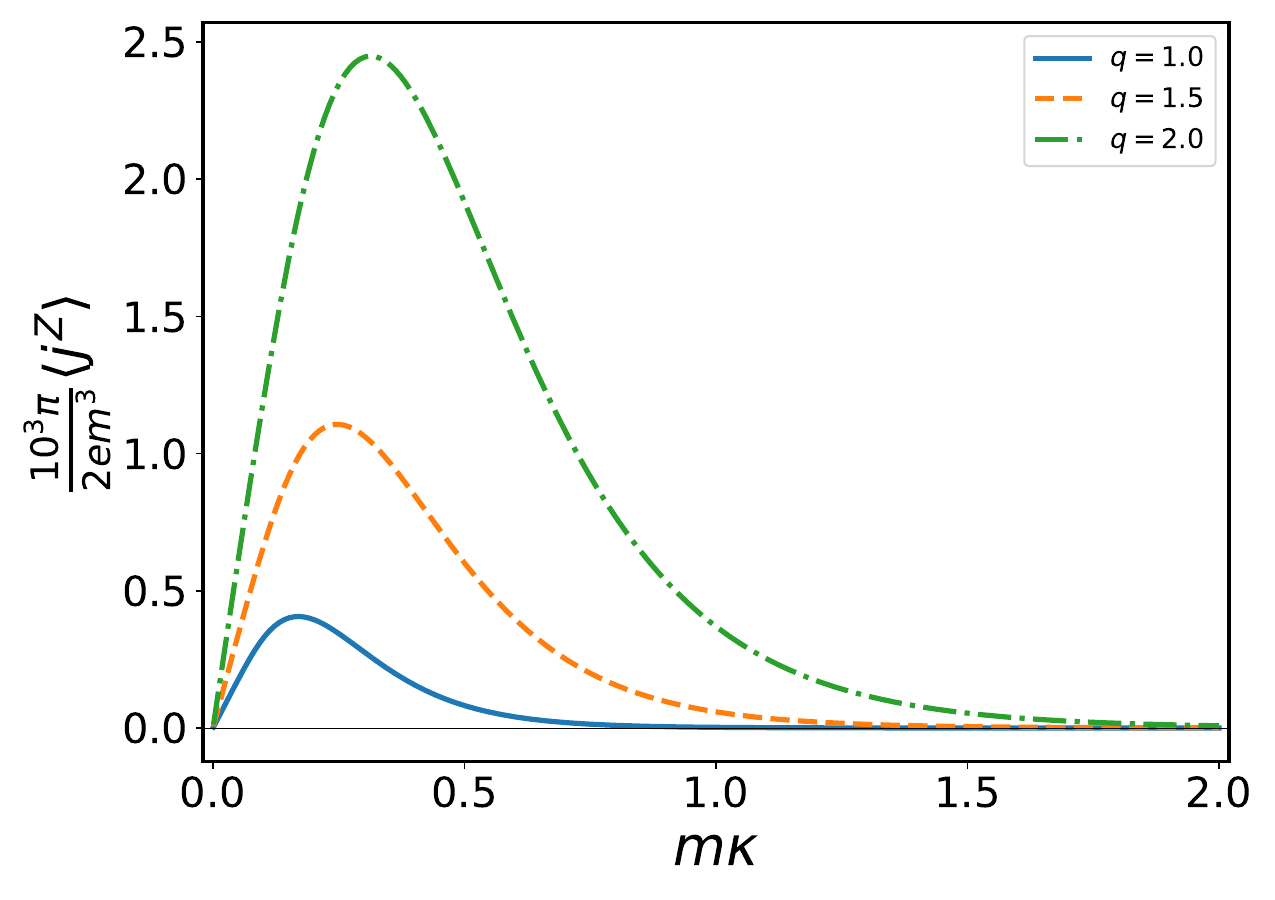}
    \includegraphics[width=0.4\textwidth]{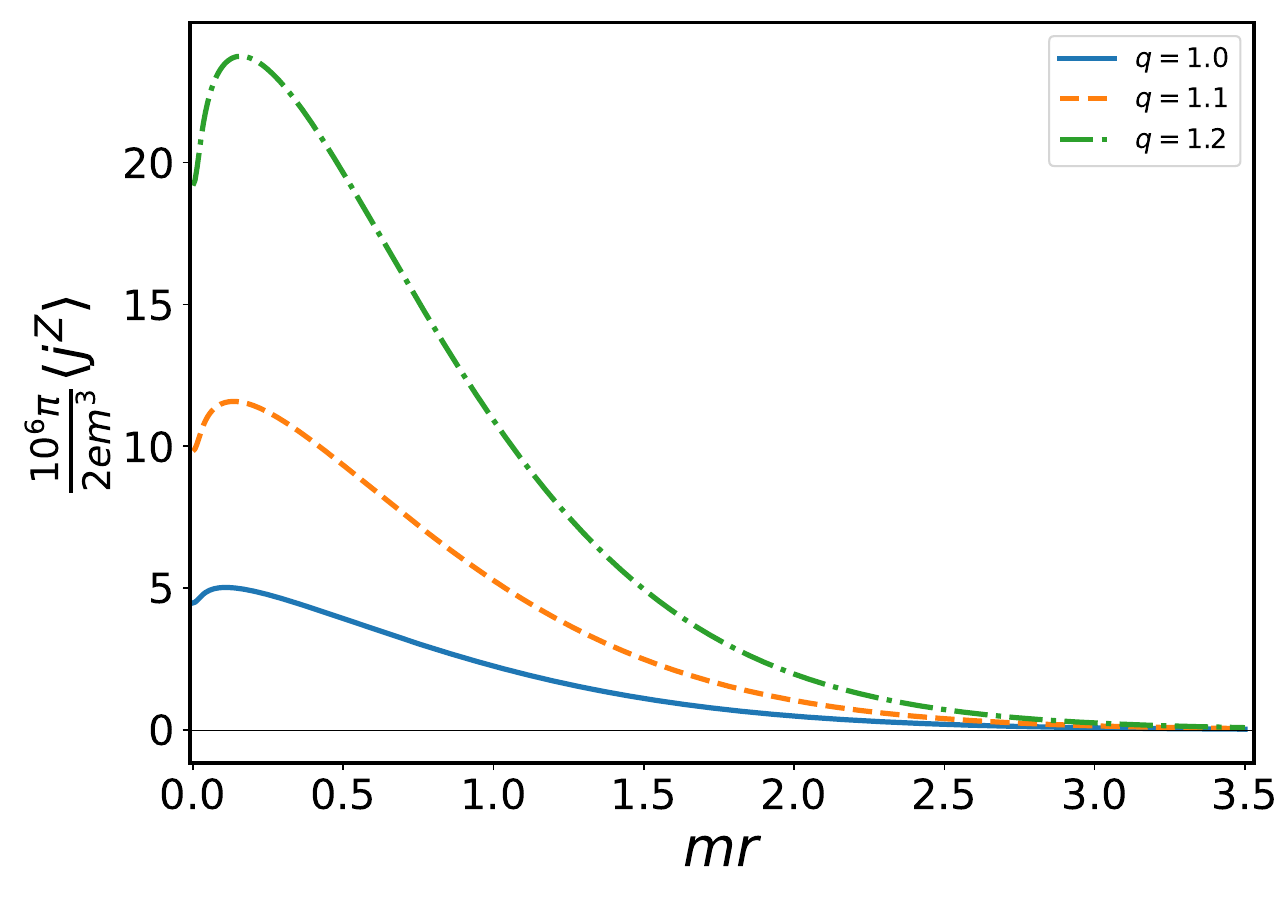}
    \caption{Plot of the induced current~\eqref{BC_def_Axial2_final_D=3} in dimensionless form as a function of $\alpha_0$ (top panels) and as a function of $m\kappa$ and $mr$ (bottom panels), for different values of $q$. In the top left panel, we set $mr=1$ and $m\kappa=0.1$, whereas in the top right panel we take $mr=m\kappa=1$. In the bottom panels, we fix $\alpha_0=0.25$, with $mr=1$ in the left panel and $m\kappa=1$ in the right panel.}
    \label{fig4}
\end{figure}
\subsubsection{Case $D=3$}
%
For the massive axial current, by specifying $D=3$ in Eq.~\eqref{BC_def_Axial2_final}, we obtain the corresponding expression in four-dimensional spacetime. Analogously to the azimuthal current, this case is also of particular physical interest, as it allows for a more direct comparison with realistic models and facilitates the analysis of the behavior of the induced current under physically relevant conditions. In this case, we have
\begin{eqnarray}
\left\langle j^Z(x)\right\rangle&=&
\frac{2\kappa em^{4}}{\pi q}\left\{\sum_{\ell=1}^{[q/2]}\ell\sin(2\ell\pi\alpha_0)f_{2}\left(m\sqrt{(p\ell)^2 + (2rs_{\ell})^2}\right)\right.\nonumber\\
&+&\left.\frac{q}{2\pi^2}\int_0^{\infty}dy\sum_{n=-\infty}^{\infty}M_n(y,\alpha_0,q)f_{2}\left(m\sqrt{(pn)^2 + (2rs_{y})^2}\right)\right\}.
\label{BC_def_Axial2_final_D=3}
\end{eqnarray}
This expression is illustrated in Fig.~\ref{fig4} in dimensionless form as a function of $\alpha_0$, $m\kappa$, and $mr$, for different values of $q$. The top panels display the periodic dependence of the induced current on $\alpha_0$. In the top left panel, we fix $mr=1$ and $m\kappa=0.1$, whereas in the top right panel we take $mr=m\kappa=1$. 
A comparison between these panels reveals a suppression in the magnitude of the induced current as $m\kappa$ increases. At the same time, the results indicate that for small but nonzero values of $\kappa$ (e.g., $m\kappa=0.1$), the current attains larger amplitudes compared to the case of larger $\kappa$. In contrast, for $\kappa=0$, the axial current vanishes identically. 
This behavior indicates the existence of an intermediate regime in which the axial current is enhanced as $\kappa$ departs from zero, before being suppressed for larger values of $m\kappa$.

In the bottom panels, where the induced current is displayed as a function of $m\kappa$ and $mr$, we fix $mr=1$ in the left panel and $m\kappa=1$ in the right panel. In both cases, we take $\alpha_0=0.25$. These plots indicate a suppression of the induced current for large values of $m\kappa$ and $mr$. At $m\kappa=0$, corresponding to the horizontal axis, the axial current vanishes, in contrast to the azimuthal component, which approaches a constant value. On the other hand, as shown in the bottom right panel, at $r=0$ the axial component assumes constant values for fixed $m\kappa$. Remarkably, no divergence arises in this case, even when both $\kappa$ and $r$ vanish.

An additional noteworthy aspect is that an approximate expression for~\eqref{BC_def_Axial2_final_D=3} can be obtained in the regime of small $m\kappa \ll 1$. In this limit, one may set $\kappa=0$ in the arguments of the Bessel functions appearing in Eq.~\eqref{BC_def_Axial2_final_D=3}, while retaining the explicit dependence on $\kappa$ in the overall prefactor. This procedure leads to a linear dependence of the axial current on the parameter $\kappa$, namely, $\left\langle j^Z(x)\right\rangle \propto \kappa$. 
For sufficiently small values of $m\kappa$, this behavior is clearly illustrated in the bottom left panel of Fig.~\ref{fig4}. This behavior is also present for a generic number of spatial dimensions $D$, both in the massive and massless cases.

For $D=3$, the massless scalar field case follows from Eqs.~\eqref{BC_def_axial_final_massless} and~\eqref{lim}, yielding the corresponding expression for the axial current in $(3+1)$-dimensional spacetime, i.e.,
\begin{eqnarray}
\left\langle j^Z(x)\right\rangle&=&
\frac{4\kappa e}{q\pi}\left\{\sum_{\ell=1}^{[q/2]}\frac{\ell\sin(2\pi \ell\alpha_0)}{\left[(p\ell)^2 + (2rs_\ell)^2\right]^{2}}\right.\nonumber\\
&+&\left.\frac{q}{2\pi^2}\int_0^{\infty}dy\sum_{n=-\infty}^{\infty}\frac{M_n(y,\alpha_0,q)}{\left[(pn)^2 + (2rs_y)^2\right]^{2}}\right\}.
\label{BC_def_axial_final_massless_D=3}
\end{eqnarray}
%
\begin{figure}
    \centering
    \includegraphics[width=0.4\textwidth]{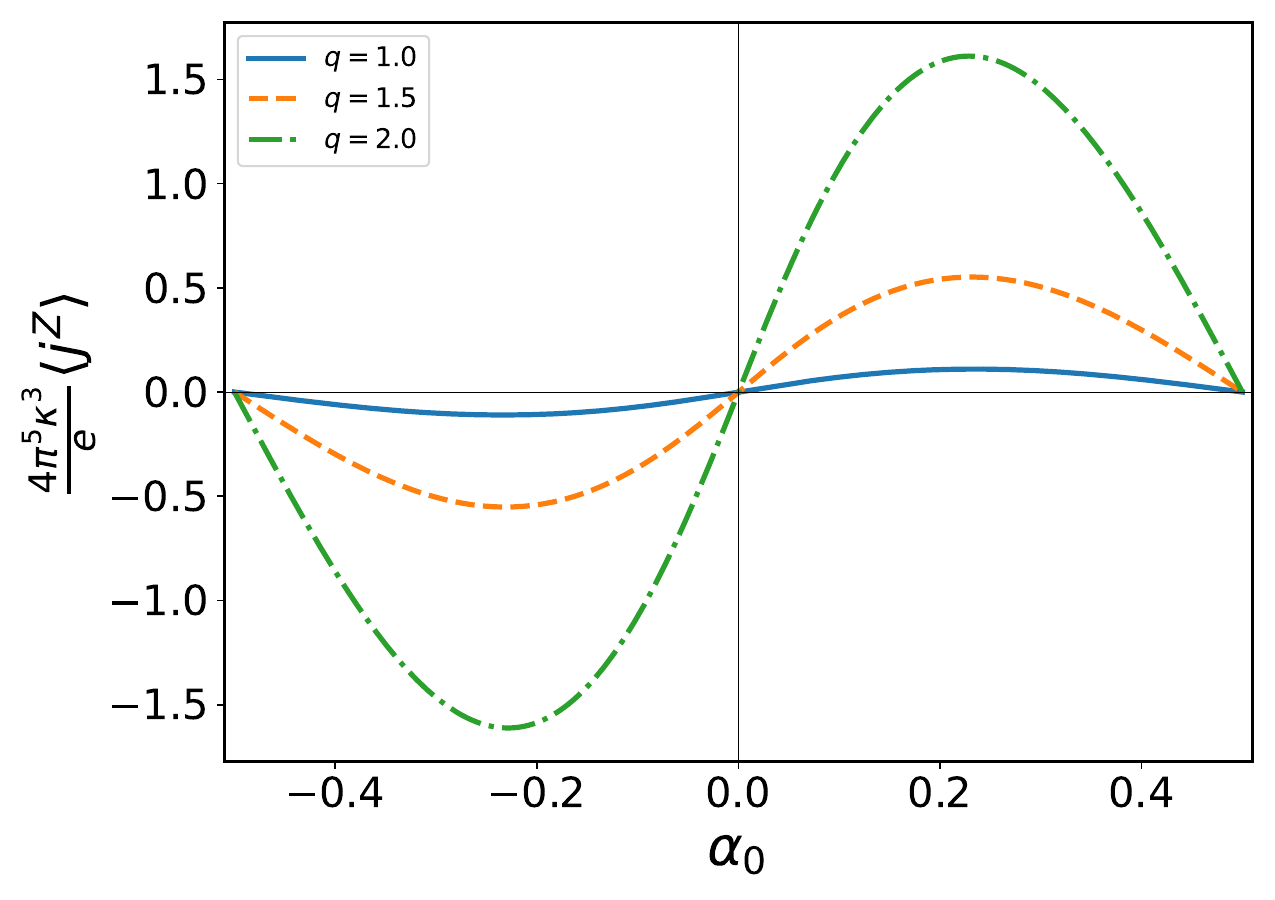}
    \includegraphics[width=0.4\textwidth]{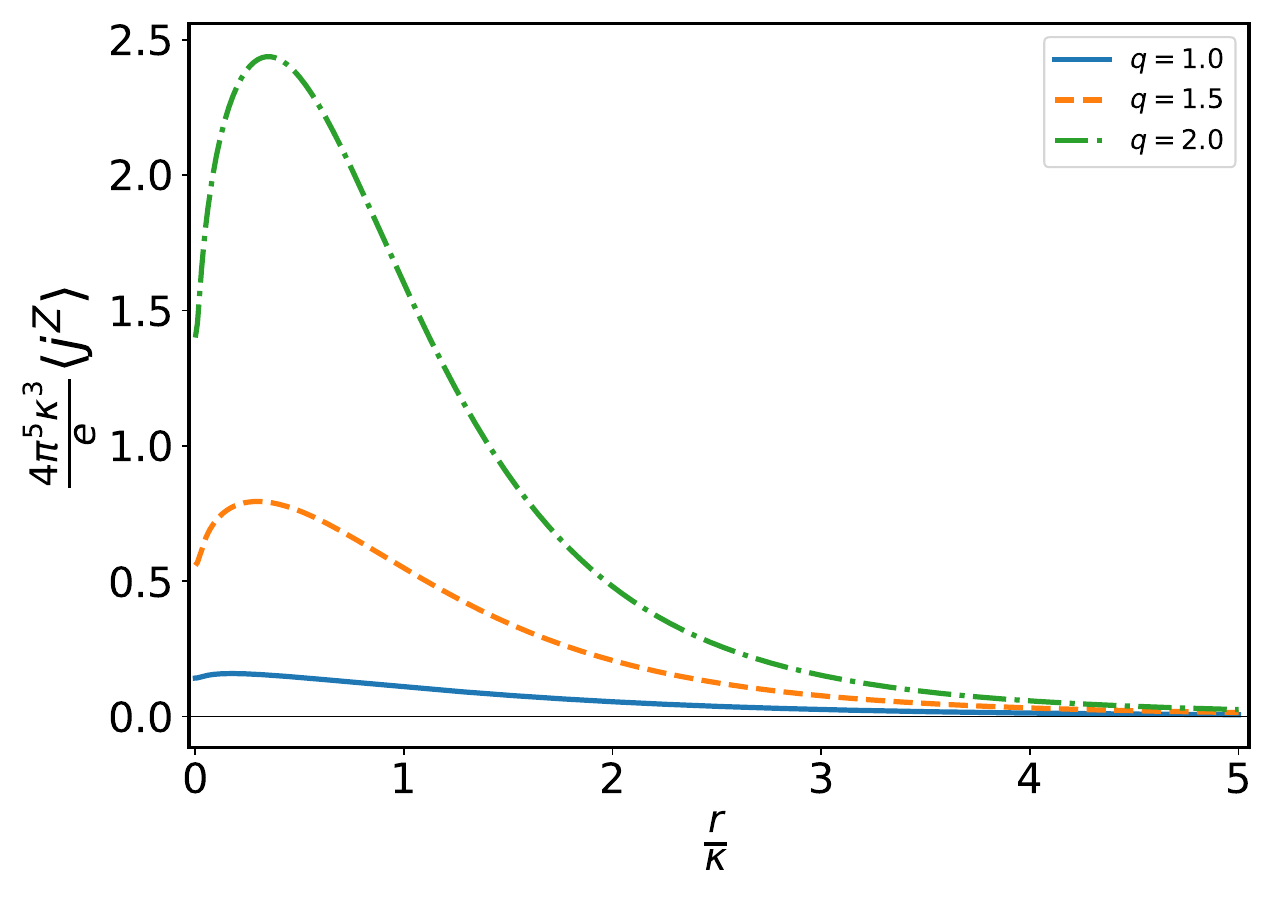}
    \caption{Plot of the induced current~\eqref{BC_def_axial_final_massless_D=3} in dimensionless form as a function of $\alpha_0$ (left panel) and of $r/\kappa$ (right panel), for different values of $q$. In the left panel, we fix $r/\kappa=1$, whereas in the right panel we take $\alpha_0=0.25$.}
 \label{fig5}
\end{figure}
This expression is displayed in Fig.~\ref{fig5} in dimensionless form as a function of $\alpha_0$ (left panel) and of $r/\kappa$ (right panel). In the left panel, we fix $r/\kappa=1$, whereas in the right panel we take $\alpha_0=0.25$. The periodic dependence of the induced current on $\alpha_0$ is again observed, along with a strong suppression for large values of $r/\kappa$. 
At $r/\kappa=0$, the current attains values consistent with the behavior $\left\langle j^{Z}(x)\right\rangle \propto \kappa^{-3}$. This corresponds to a power-law dependence distinct from that of the massless azimuthal component, indicating that the axial current decays more slowly with $\kappa$ at the origin.

Overall, the results obtained in this section highlight the combined role of topology and gauge fields in determining the behavior of the induced currents. The dependence on the parameter $\alpha_0$ reflects the underlying Aharonov--Bohm-type effect, while the suppression for large values of $r/\kappa$ reveals the interplay between the radial distance and the screw dislocation parameter. These features arise consistently in both the massive and massless cases, for the azimuthal and axial current densities, thereby reinforcing the robustness of the physical effects discussed.
\section{Conclusions and Remarks}
\label{secV}

In this work, we have investigated the vacuum-induced current density associated with a charged scalar field in a $(D+1)$-dimensional cosmic dispiration spacetime threaded by a magnetic flux. This background combines a conical defect, characterized by the parameter $q$, with a screw dislocation encoded in the parameter $\kappa$, providing a nontrivial global structure despite being locally flat.

By constructing a complete set of normalized solutions of the Klein--Gordon equation, we have obtained the positive-frequency Wightman function and used it to evaluate the vacuum expectation value of the current density. Our analysis shows that, in addition to the azimuthal component, $\langle j^\phi(x) \rangle$, which describes a persistent vacuum current circulating around the defect, a nonvanishing axial component, $\langle j^Z(x) \rangle$, is also induced. The latter arises as a direct consequence of the helical structure of the spacetime, reflecting the nontrivial coupling between the angular and longitudinal coordinates.

A central result of this work is the derivation of closed expressions for both components of the induced current density, for massive and massless fields in arbitrary dimensions. These expressions clearly reveal the interplay between geometry, topology, and gauge fields in determining the vacuum polarization effects.

We have shown that both components of the induced current are periodic functions of the magnetic flux, depending only on its fractional part, $\alpha_0$. This behavior reflects the Aharonov--Bohm nature of the effect, where physical observables are sensitive to the gauge potential even in regions where the corresponding field strength vanishes. In particular, both currents vanish for $\alpha_0=0$ and $\alpha_0=1/2$.

An important feature of our results is the role played by the screw dislocation parameter $\kappa$. While the azimuthal current persists even in the absence of the dislocation, the axial component is entirely induced by it and vanishes for $\kappa=0$. Moreover, in contrast to the case of a pure cosmic string, the presence of $\kappa$ leads to a finite and well-defined current density at the position of the defect, $r=0$. In this sense, the helical torsion associated with the dispiration acts as an effective regulator. In the massless case, at the origin, both the azimuthal and axial currents exhibits a power-law dependence on $\kappa$, highlighting its role as the relevant scale controlling the behavior of vacuum fluctuations.

We have also analyzed the asymptotic behavior of the induced currents. For large values of the parameters $mr$ and $m\kappa$, both components are exponentially suppressed, reflecting the short-range nature of vacuum polarization effects for massive fields. In contrast, in the massless limit, the currents display long-range behavior governed by inverse power laws, making the topological effects more pronounced.

Our results reduce, in the limit $\kappa \to 0$, to the known expressions for the induced current in a cosmic string spacetime with a magnetic flux, providing a nontrivial consistency check of the formalism. This demonstrates that the present analysis extends previous results by systematically incorporating the effects of helical torsion due to the screw dislocation.

From a physical perspective, the induced current densities can be interpreted as manifestations of vacuum polarization in a nontrivial topological and gauge background. In semiclassical approaches, such currents act as sources in Maxwell's equations and may lead to observable electromagnetic effects. Moreover, the sensitivity of the currents to global properties of the spacetime reinforces the inherently nonlocal character of quantum vacuum phenomena.

Finally, the results presented here may also find applications in analogous condensed matter systems, where defects such as disclinations and screw dislocations can be effectively described by similar geometrical frameworks. In this context, the induced currents analyzed in this work may correspond to measurable persistent currents, providing a bridge between high-energy physics and condensed matter realizations.

Possible extensions of this work include the investigation of fermionic fields, finite temperature effects, and the inclusion of non-minimal coupling between the scalar field and curvature. Another interesting direction would be the analysis of backreaction effects, where the induced currents are incorporated as sources in the dynamical equations for the gauge field.

\acknowledgments

The author thanks Eugênio R.~Bezerra de Mello for valuable discussions. This work was partially supported by the Brazilian National Council for Scientific and Technological Development (CNPq) under Grant No.~308049/2023-3.

\appendix

\section{Expression for the function $\mathcal{J}(w)$}
\label{Ap_a}
Our goal in this appendix is to derive a more convenient representation for the function defined in Eq.~\eqref{J_fun}, which, for convenience, we reproduce here. That is,
\begin{equation}
\mathcal{J}_{\phi}(w)=\int_{-\infty}^{\infty}\frac{qdh}{\kappa}e^{-\frac{q^2r^2 h^2}{2w\kappa^2}}
\sum_{n=-\infty}^{\infty}\beta_{\sigma}
I_{\beta_{\sigma}}(w),
\label{J_fun_A}
\end{equation}
where $\beta_{\sigma}=q|n-h+\alpha_0|$, and $I_{\beta_{\sigma}}(w)$ denotes the modified Bessel function of the first kind.

By making use of the delta-function property $\int dz\,\delta(z-z_0)g(z)=g(z_0)$, together with the Poisson summation formula,
\begin{equation}
2\pi\sum_{n=-\infty}^{\infty}\delta(b - 2\pi n)=\sum_{n=-\infty}^{\infty}e^{ib n},
\label{iden}
\end{equation}
we obtain, upon setting $b = 2\pi(x + h - \alpha_0)$,
\begin{eqnarray}
\mathcal{J}_{\phi}(w)=\sum_{n=-\infty}^{\infty}\int_{-\infty}^{\infty}\frac{qdh}{\kappa}e^{-\frac{q^2r^2 h^2}{2w\kappa^2} + i2\pi hn}
\int_{-\infty}^{\infty}dx e^{i2\pi n(x - \alpha_0)}qx
I_{q|x|}(w).
\label{A.7.2}
\end{eqnarray}

We can now carry out the integration over $h$ and decompose the integral over $x$ into two parts. This yields
\begin{equation}
\mathcal{J}_{\phi}(w)=-\frac{2i(2\pi w)^{\frac{1}{2}}}{r}
\sum_{n=-\infty}^{\infty}\sin(2\pi n\alpha_0)e^{-\frac{p^2n^2w}{2r^2}}\int_{0}^{\infty}dxe^{i2\pi nx} qx
I_{qx}(w),
\label{A.7.3}
\end{equation}
where $p = 2\pi\kappa/q$ represents the helical pitch of the cosmic dispiration.

We next make use of the recurrence relation~\cite{abramowitz1965handbook}
\begin{eqnarray}
q x\, I_{q x}(w) = - w \frac{d}{d w} I_{q x}(w) + w I_{q x - 1}(w),
\label{recurrence_bessel}
\end{eqnarray}
together with the integral representation of the modified Bessel function,
\begin{equation}
I_{\gamma}(w)=
\frac{1}{\pi}
\int_{0}^{\pi}
e^{w\cos\theta}\cos(\gamma\theta)d\theta
-
\frac{\sin(\pi\gamma)}{\pi}
\int_{0}^{\infty}
e^{-w\cosh y-\gamma y}dy,
\label{A.8}
\end{equation}
to show that
\begin{eqnarray}
I_{qx -1 }(w)&=&\frac{d}{dw}I_{qx}(w)\nonumber\\
&+&\frac{1}{\pi}\int_0^{\pi}d\theta e^{w\cos\theta}\sin\theta\sin(qx\theta)\nonumber\\
&+&\frac{\sin(q\pi x)}{\pi}\int_0^{\infty}dye^{-w\cosh y}\sinh y e^{-qxy}.
\label{result1}
\end{eqnarray}
Consequently, upon substituting Eqs.~\eqref{recurrence_bessel} and~\eqref{result1} into Eq.~\eqref{A.7.3}, we can perform the integration over $x$, yielding
\begin{eqnarray}
\mathcal{J}_{\phi}(w)&=&\frac{2(2\pi)^{\frac{1}{2}}w^{\frac{3}{2}}}{qr}
\sum_{\ell=1}^{[q/2]}e^{-\frac{p^2\ell^2w}{2r^2}}\sin(2\pi\ell/q)\sin(2\pi \ell\alpha_0)e^{w\cos(2\pi\ell/q)}\nonumber\\
&+&\frac{(2\pi)^{\frac{1}{2}}w^{\frac{3}{2}}}{\pi^2 r}\int_0^{\infty}dy\, e^{-w\cosh y}\sinh y\sum_{n=-\infty}^{\infty}e^{-\frac{p^2n^2w}{2r^2}}S_n(y,\alpha_0,q),
\label{J_fun_text}
\end{eqnarray}
where we have defined
\begin{eqnarray}
S_n(y,\alpha_0,q)=\frac{\left(\frac{qy}{2\pi}\right)\sin(2\pi n\alpha_0)}{\left(n - \frac{q}{2}\right)^2 + \left(\frac{qy}{2\pi}\right)^2}.
\label{sum1}
\end{eqnarray}
The sum over $\ell$ in Eq.~\eqref{J_fun_text} originally obeys the restriction
\begin{eqnarray}
-\frac{q}{2} < \ell < \frac{q}{2},
\label{restriction}
\end{eqnarray}
which can be decomposed into two equal contributions ranging from $\ell=1$ to $\ell=[q/2]$, where $[q/2]$ denotes the integer part of $q/2$. In the special case $q = 2g$, with $g$ an integer, the term corresponding to $\ell = q/2$ must be taken with a weight factor of $1/2$. Moreover, the first term on the r.h.s. exists only for $q \geq 2$; otherwise, it is absent. With this result at hand, we can further develop the azimuthal current density given in Eq.~\eqref{BC_def_azimuthal2}, leading to the closed-form expressions in Eqs.~\eqref{BC_def_azimuthal_final} and~\eqref{BC_def_azimuthal_final_massless}.
\section{Expression for the function $\mathcal{J}_Z(w)$}
\label{Ap_axial}
%
Let us now work out an expression for the function $\mathcal{J}_Z(w)$, defined in Eq.~\eqref{J_fun_Z}. We start by
writing it here again, i.e.,
\begin{equation}
\mathcal{J}_Z(w)=\frac{q^2}{\kappa^2}\int_{-\infty}^{\infty}hdh
e^{-\frac{q^2 r^2 h^2}{2w\kappa^2}}
\sum_{n=-\infty}^{\infty}
I_{\beta_{\sigma}}(w).
\label{J_fun_Z_App}
\end{equation}

Upon making use of the Poisson summation formula in Eq.~\eqref{iden}, with $b = 2\pi(x + h - \alpha_0)$, and performing the integration
over $h$, we have
\begin{eqnarray}
\mathcal{J}_Z(w)&=&\frac{i\kappa(2\pi w)^{\frac{3}{2}}}{qr^3}\sum_{n=-\infty}^{\infty}ne^{-\frac{p^2n^2w}{2r^2}}
\int_{-\infty}^{\infty}dx e^{i2\pi n(x - \alpha_0)}
I_{q|x|}(w)\nonumber\\
&=&\frac{i\kappa(2\pi w)^{\frac{3}{2}}}{qr^3}(T_1 + T_2),
\label{J_axial}
\end{eqnarray}
where using the integral representation~\eqref{A.8} we can separate the expression above into two terms, $T_1$ e $T_2$. The contribution $T_1$ comes from the first term on the r.h.s. of~\eqref{A.8}, that is, 
\begin{eqnarray}
T_1 &=& \frac{1}{\pi}\sum_{n=-\infty}^{\infty}ne^{-\frac{p^2n^2w}{2r^2}}\int_{-\infty}^{\infty}dx e^{i2\pi n(x - \alpha_0)}\int_0^{\pi}d\theta e^{w\cos\theta}\cos(qx\theta)\nonumber\\
&=& \frac{1}{q}\sum_{\ell} \ell e^{-\frac{p^2\ell^2w}{2r^2}}e^{w\cos(2\pi\ell/q) }e^{-i2\pi\ell\alpha_0},
\label{termT_1}
\end{eqnarray}
where $\ell$ obeys the restriction in Eq.~\eqref{restriction}.

The second contribution, $T_2$, on the other hand, comes from the second term on the r.h.s. of~\eqref{A.8}. This can be write as
\begin{eqnarray}
T_2 &=&- \frac{1}{\pi}\sum_{n=-\infty}^{\infty}ne^{-\frac{p^2n^2w}{2r^2}}\int_{-\infty}^{\infty}dx\sin(q|x|\pi) e^{i2\pi n(x - \alpha_0)}\int_0^{\infty}dy e^{-w\cosh y-q|x|y}\nonumber\\
&=&  \frac{2i}{\pi}\sum_{n=-\infty}^{\infty}n\sin(2\pi n\alpha_0)e^{-\frac{p^2n^2w}{2r^2}}\int_0^{\infty}dy e^{-w\cosh y}\int_{0}^{\infty}dx\sin(qx\pi) e^{i2\pi n x - qxy}\nonumber\\
&=& - \frac{i}{\pi^2}\int_0^{\infty}dy e^{-w\cosh y}\sum_{n=-\infty}^{\infty}e^{-\frac{p^2n^2w}{2r^2}}M_n(y, \alpha_0, q).
\label{termT_2}
\end{eqnarray}
where
\begin{eqnarray}
M_n(y,\alpha_0,q)=\frac{\left(n - \frac{q}{2}\right)n\sin(2\pi n\alpha_0)}{\left(n - \frac{q}{2}\right)^2 + \left(\frac{qy}{2\pi}\right)^2}.
\label{sum_M}
\end{eqnarray}
Consequently, from Eqs.~\eqref{termT_1}, \eqref{termT_2} and~\eqref{J_axial}, we obtain
\begin{eqnarray}
\mathcal{J}_Z(w)&=&\frac{2\kappa(2\pi w)^{\frac{3}{2}}}{q^2r^3}\sum_{\ell=1}^{[q/2]} \ell\sin(2\pi\ell\alpha_0) e^{-\frac{p^2\ell^2w}{2r^2}}e^{w\cos(2\pi\ell/q)}\nonumber\\
&+&\frac{\kappa(2\pi w)^{\frac{3}{2}}}{qr^3\pi^2}\int_0^{\infty}dy e^{-w\cosh y}\sum_{n=-\infty}^{\infty}e^{-\frac{p^2n^2w}{2r^2}}M_n(y, \alpha_0, q).
\label{J_axial2}
\end{eqnarray}
Note that the same discussion made below Eq.~\eqref{restriction} for the sum in $\ell$ also applies for the expression above. 

Unfortunately, the sum over $n$ in Eqs.~\eqref{J_fun_text} and~\eqref{J_axial2} does not admit a closed-form analytical expression. However, upon setting $\kappa = 0$ in Eq.~\eqref{J_fun_text}, the series over $n$ of $S_n(y,\alpha_0,q)$ can be evaluated analytically, which will be demonstrated next.  For a null $\kappa$ the expression in Eq.~\eqref{J_axial2} vanishes. This shows that when $\phi$ decouples from $Z$, the only nonzero component of the induced current is the azimuthal one.

\section{Evaluation of the series over $n$ for the function $S_n(y,\alpha_0,q)$}
\label{Ap_series}
As stated before, the sum over $n$ in Eq.~\eqref{J_fun_text} does not admit a closed-form expression. However, in the particular case $\kappa = 0$, the series can be evaluated explicitly. In this appendix, we focus on deriving an analytical expression for the resulting series, which we write as
\begin{equation}
\sum_{n=-\infty}^{\infty}
S_n(y,\alpha_0,q)
=
\sum_{n=-\infty}^{\infty}
\frac{A\,\sin(2\pi n \alpha_0)}
{(n-a)^2 + A^2},
\label{series_start}
\end{equation}
where we have defined
\begin{equation}
A = \frac{qy}{2\pi}, \qquad a = \frac{q}{2}.
\label{B2}
\end{equation}

We can now express Eq.~\eqref{series_start} in terms of the imaginary part of the expression below, i.e.,
\begin{equation}
\sum_{n=-\infty}^{\infty} S_n
=
A\,\mathrm{Im}
\left[
\sum_{n=-\infty}^{\infty}
\frac{e^{i2\pi n\alpha_0}}{(n-a)^2 + A^2}
\right].
\label{imag_form}
\end{equation}
This representation allows us to apply the Poisson summation formula,
\begin{equation}
\sum_{n=-\infty}^{\infty} f(n)
=
\sum_{k=-\infty}^{\infty} \tilde f(k),
\label{PSF}
\end{equation}
where
\begin{equation}
\tilde f(k)
=
\int_{-\infty}^{\infty}
dx\, f(x)e^{-i2\pi k x},
\label{FT}
\end{equation}
is the Fourier transform of the function $f(x)$.

From Eq.~\eqref{imag_form}, we then define
\begin{equation}
f(x) = \frac{e^{i2\pi \alpha_0 x}}{(x-a)^2 + A^2}.
\end{equation}
Consequently, the integral in Eq.~\eqref{FT} becomes
\begin{equation}
\tilde f(k)
=
\int_{-\infty}^{\infty}
\frac{e^{i2\pi (\alpha_0-k)x}}{(x-a)^2 + A^2} \, dx.
\end{equation}

By performing the change of variable $u = x-a$, we obtain
\begin{equation}
\tilde f(k)
=
e^{i2\pi (\alpha_0-k)a}
\int_{-\infty}^{\infty}
\frac{e^{i2\pi (\alpha_0-k)u}}{u^2 + A^2} \, du.
\label{integral1}
\end{equation}

Using the standard integral
\begin{equation}
\int_{-\infty}^{\infty}
\frac{e^{i\lambda u}}{u^2 + A^2} \, du
=
\frac{\pi}{A} e^{-A|\lambda|},
\end{equation}
we can evaluate the integral in Eq.~\eqref{integral1}, yielding
\begin{equation}
\tilde f(k)
=
\frac{\pi}{A}
e^{i2\pi (\alpha_0-k)a}
e^{-2\pi A|\alpha_0-k|}.
\end{equation}

Thus, from Eq.~\eqref{PSF}, we obtain
\begin{eqnarray}
\sum_{n=-\infty}^{\infty}
\frac{e^{i2\pi n\alpha_0}}{(n-a)^2 + A^2}
&=&
\frac{\pi}{A}
\sum_{k=-\infty}^{\infty}
e^{i2\pi (\alpha_0-k)a}
e^{-2\pi A|\alpha_0 - k|} \nonumber\\
&\equiv& \frac{\pi}{A} S,
\label{B11}
\end{eqnarray}

In the present analysis, we assume $-1/2 < \alpha_0 < 1/2$. For $\alpha_0 > 0$, the sum over $k$ can be decomposed as
\begin{equation}
S = S_1 + S_2,
\label{B12}
\end{equation}
with
\begin{eqnarray}
S_1 &=& \sum_{k\leq 0}
e^{i2\pi a(\alpha_0-k)}
e^{-2\pi A(\alpha_0 - k)},\\
S_2 &=& \sum_{k\geq 1}
e^{i2\pi a(\alpha_0 - k)}
e^{-2\pi A(k - \alpha_0)}.
\end{eqnarray}

By defining $k=-m$ in $S_1$ and $k=m$ in $S_2$, we obtain
\begin{eqnarray}
S_1 &=& \sum_{m=0}^{\infty}
e^{i2\pi a(\alpha_0 + m)}
e^{-2\pi A(\alpha_0 + m)},\\
S_2 &=& \sum_{m=1}^{\infty}
e^{i2\pi a(\alpha_0 - m)}
e^{-2\pi A(m - \alpha_0)}.
\end{eqnarray}

The sums over $m$ above are geometric series and can be written as
\begin{equation}
S_1 =
\frac{
e^{-2\pi A\alpha_0} e^{i2\pi a\alpha_0}
}{
1 - e^{-2\pi A} e^{i2\pi a}
},
\end{equation}
and
\begin{equation} 
S_2 = \frac{ e^{2\pi A\alpha_0} e^{i2\pi a\alpha_0} }{ e^{2\pi A} e^{i2\pi a} - 1 }. 
\end{equation}

Combining the above expressions, we obtain
\begin{equation}
S_1 + S_2 =
\frac{
\sinh(2\pi A\alpha_0)e^{-i2\pi a(1-\alpha_0)}
+
\sinh[2\pi A(1-\alpha_0)]e^{i2\pi a\alpha_0}
}{
\cosh(2\pi A) - \cos(2\pi a)
}.
\label{B19}
\end{equation}
Substituting Eq.~\eqref{B19} into~\eqref{B12}, and then into~\eqref{B11}, and finally into~\eqref{imag_form}, after taking the imaginary part, we find
\begin{equation}
\sum_{n=-\infty}^{\infty} S_n
=\pi
\frac{
\sin(2\pi a \alpha_0)\sinh[2\pi A (1-\alpha_0)]
-
\sin[2\pi a (1-\alpha_0)]\sinh(2\pi A \alpha_0)
}{
\cosh(2\pi A) - \cos(2\pi a)
}.
\label{B20}
\end{equation}
Note that, if $\alpha_0 < 0$ in Eq.~\eqref{B11}, the resulting expression differs from the one above only by an overall minus sign. Therefore, by using Eq.~\eqref{B2}, we can rewrite the above result in a form that is valid for both positive and negative values of $\alpha_0$ as
\begin{equation}
\sum_{n=-\infty}^{\infty} S_n(y,\alpha_0,q)
=\pi
\frac{
\sin(q\pi \alpha_0)\sinh[(1-|\alpha_0|)qy]
-
\sin[(1-|\alpha_0|)q\pi]\sinh(qy\alpha_0)
}{
\cosh(qy) - \cos(q\pi)
}.
\label{B21}
\end{equation}
This expression exactly reproduces the known results reported in Refs.~\cite{Braganca:2014qma, BezerradeMello:2014phm} in the cosmic string limit, $\kappa \to 0$, of Eqs.~\eqref{BC_def_azimuthal_final} and~\eqref{BC_def_azimuthal_final_massless}.


\end{document}